\documentstyle[prd,aps]{revtex}
\begin{document}
\draft
\title{Neutrino-electron scattering in dense magnetized plasma}

\author{V.G.~Bezchastnov}
\address{A.F.~Ioffe Physical-Technical Institute,
         194021 St.Petersburg, Russia}
\author{P.~Haensel}
\address{N.~Copernicus Astronomical Center,
         Polish Academy of Sciences, Bartycka 18,
         00-716 Warszawa, Poland}
\date{\today}
\maketitle

\begin{abstract}
We derive general expressions for the cross section
of neutrino scattering on electrons in dense, hot
stellar matter, in the presence of  strong magnetic fields.
Numerical calculations of the scattering cross sections at various
densities, temperatures and magnetic fields, are performed.
Strong, quantizing magnetic fields modify significantly the
angular and energy dependence of the scattering cross section.
\end{abstract}

\vskip 0.5cm
\centerline{\sl to be published in Physical Review D (September, 1996)}
\vskip 0.5cm


\section*{Introduction}
\label{Intro}
Neutrinos play a crucial role in the gravitational collapse
of massive, evolved stellar cores, which is thought to be
at the origin
of Type II supernova explosions. During the infall phase,
when the collapsing stellar core increases its central
density  from the initial
$\sim 10^{10}~{\rm g~cm^{-3}}$
to  the supernuclear one, exceeding
$10^{14}~{\rm g~cm^{-3}}$,
a large number of electron neutrinos is produced, as a result of
electron captures on protons (both free protons and those bound
in nuclei). The role of neutrinos of other
flavors is, during the infall phase, negligible.
The fate of electron neutrinos is
determined by the opacity of the collapsing core, which, in turn,
is determined by their interaction with the dense, hot medium.
The main source of neutrino opacity is the elastic scattering
on nuclei, which leads to neutrino trapping at
$\rho\sim 10^{12} - 10^{13}~{\rm g~cm^{-3}}$.
Simple estimates suggest that
scattering of neutrinos on degenerate electrons is not
important as a source of opacity, therefore this
process was not included in the earlier collapse simulations
(see, e.g., \cite{coo88} and the references therein).
However, as pointed out in \cite{bru85,bru88,myr88},
and confirmed recently in \cite{mez93},
neutrino-electron scattering is a
significant source of matter entropy increase as well as of core
deleptonization, and therefore its
inclusion is important for the reliable
determination of the fate of the collapsing core.

The special role of the neutrino-electron scattering (NES),
\begin{equation}
    \nu + e \rightarrow \nu + e,
\label{eq:Reaction}
\end{equation}
results from the fact that it can be accompanied
by a significant energy
transfer, in contrast to the case of neutrino-nucleus or
neutrino-nucleus scattering, which are to a very good
approximation elastic (conservative). Electron capture on free
protons (which is a main source of electron neutrinos)
produces neutrinos, which are ``hotter" than the matter
(i.e., their mean energy is significantly
larger than that corresponding to matter
temperature). As the neutrinos flow from denser to less dense
layers of the collapsing core, this excess is even higher than the
local one. Energy is transferred from neutrinos to matter mostly
by NES, which tends to equilibrate
neutrinos and matter. Because on average neutrinos loose energy
in the NES, this process decreases the mean
neutrino energy, and in consequence, increases its mean free
path (the neutrino mean free path scales as the
inverse square of the neutrino energy):
down-scattered neutrinos escape more easily from
the collapsing core. On the other hand, neutrino
down-scattering heats the matter
(increases the matter entropy),  which increases the free proton
fraction, and this accelerates electron captures. Both
effects lead to significantly higher deleptonization of the
collapsing core (as compared to the case with NES turned off,
see \cite{mez93}),
which turns out to be crucial for the
energetics of the post-bounce shock, and to the eventual success
or failure of the shock to produce  an explosion of the collapsing
star.

While observations tell that young neutron stars possess a
very strong magnetic field (external magnetic field of radio
pulsars $B \sim 10^{12}-10^{13}~$G,
the internal magnetic field can be
significantly higher, e.g., $B \sim 10^{15}$~G), the mechanism
of its formation is still a matter of debate. Even less is known
about the role of the magnetic field during
collapse. The initial (primordial) field of the
collapsing core could be greatly amplified due to the huge
compression of  the matter of very high conductivity. Superstrong
magnetic fields  ($\sim 10^{15}~$G) could also
be generated in collapsing and rotating cores
(\cite{leb70,bis76,mue79}).
If present, such strong magnetic fields
could influence  NES by modifying
the motion of electrons in the hot dense plasma. This could lead to
anisotropies in the scattering process, implying possible
anisotropies and asymmetries in the collapse and explosion.

The problem of NES in a hot, dense
plasma in the absence of a magnetic field was studied by numerous
authors (see, e.g., \cite{yue76,sch82}).
In the present paper we calculate
the cross section of this process in the presence of a strong
magnetic field.
The elementary process we consider here is related to two other
weak interaction processes in a hot, dense, magnetized plasma
which were considered recently in a paper co-authored by one of the
authors (\cite{kam92}): neutrino-pair synchrotron
radiation by electrons,
$e^-\longrightarrow e^- +\nu + \bar\nu$,
and the neutrino-pair emission from the electron-positron
annihilation, $e^- + e^+ \longrightarrow \nu + \bar\nu$.
The formulae obtained in the present paper complete thus the
general study of weak interaction processes involving electrons,
positrons, and neutrinos in hot, dense, magnetized plasma.

In Sec.~\ref{Sect.1} we briefly present our formalism.
The exact calculation of the  NES cross
section in magnetized plasma is described in Sec.~\ref{Sect.2}.
In Sec.~\ref{Sect.3} we show, using the quasi-classical
approximation, convergence of our result to the zero-field limit.
Numerical results, obtained for various combinations of hot plasma
parameters, neutrino energies, and magnetic field strengths, are
presented in Sec.~\ref{Sect.4}. Sec.~\ref{Sect.5} contains a
discussion of our results and the conclusion. Formulae for the
transition amplitude are collected in Appendix~A.
In Appendix~B we rederive a compact expression for the
cross section in absence of a magnetic field.In Appendix~C
we give an alternative form of the cross section at $B=0$, which
proves to be  useful in our discussion of the limiting behavior of
the cross section at low magnetic field strengths.
%
\section{General formalism}
\label{Sect.1}
Let us consider NES (\ref{eq:Reaction})
in a strong uniform magnetic field {\bf B} directed along
the $z$-axis. We will use the standard  Weinberg - Salam - Glashow
weak interaction theory. Our results will
be valid for all neutrino flavors. Since typical energies
of neutrinos, relevant for the infall phase of gravitational
collapse, are greater than a few MeV, we can
adopt the approximation of massless neutrinos.
On the other hand, these typical energies are
much smaller than the $m_{\rm W} c^2 \sim 70$ GeV, the rest-mass
energy of the intermediate bosons. Therefore
the process of scattering (\ref{eq:Reaction}) can be described by a
simplest ``four-tail" diagram.

We use exact wave functions of relativistic
electrons in a magnetic field, with  the Landau gauge
of the vector potential {\bf A} ($A_x =-yB$, $A_y=A_z=0$).
We specify the quantum states of electrons by
four quantum numbers: $p_z$, $p_x$, $n$, and $s$
(see, e.g., \cite{kam81}).
Here $p_z$ is the electron momentum along {\bf B},
$p_x$ determines the $y$-coordinate of the electron
guiding center $y=y_0= -cp_x/(eB)$, $n=0,1,\ldots$ enumerates
the Landau energy levels, and $s$ denotes
the electron helicity
($s=-{\rm sgn}(p_z)$ for $n=0$ and $s=\pm 1$ for $n>0$).
We will mostly use the units
$ k_{\rm B} = \hbar = c = 1 $,
expressing momenta in units of $m_ec$, energies (and
temperature)
in $m_ec^2$,
and magnetic fields in units of $B_0=m_e^2c^3/(\hbar e)
\approx 4.414 \times 10^{13}$~G. Then the electron
energy is
\begin{equation}
     \varepsilon = \varepsilon_n(p_z) = \sqrt{1 + 2nb + p_z^2},
\label{eq:Energy}
\end{equation}
where $b=B/B_0$ is the dimensionless magnetic field.

Let us denote the initial state in process (\ref{eq:Reaction}),
corresponding to an electron state
$ | p_z, p_x, n, s \rangle $ and
incident neutrino momentum  {\bf k},
by  $ | {\rm i} \rangle $.
The final state will be denoted by
$ | {\rm f} \rangle = | p'_z, p'_x, n', s', {\bf k}' \rangle $.
The electron and neutrino energies, corresponding to
$ | {\rm i} \rangle $ and $ | {\rm f} \rangle $ states,
are $ \varepsilon $, $\omega$, and $ \varepsilon' $, $\omega'$,
respectively. The probability of the
transition $ | {\rm i} \rangle \rightarrow | {\rm f} \rangle $,
per unit time, is given by the Fermi Golden Rule,
%
\begin{equation}
{\rm d} w_{\rm f\,i} =
2 \pi \delta \left( \varepsilon' + \omega'
                  - \varepsilon - \omega \right)
| V_{\rm f\,i} |^2 {\rm d} \rho_{\rm f},
\label{eq:GoldenRule}
\end{equation}
where
%
\begin{equation}
{\rm d} \rho_{\rm f} = \frac{L_x}{2\pi} \, {\rm d} p'_x \,
                       \frac{L_z}{2\pi} \, {\rm d} p'_z \,
                       \frac{V}{(2\pi)^3} \, {\rm d} {\bf k}'
\label{eq:DensityStates}
\end{equation}
is the density of states of particles after scattering
($L_z$ and $L_x$ are the normalization lengths, and
$V$ is the normalization volume), and $V_{\rm f\,i}$ is the matrix
element, calculated for the ``four-tail" diagram of process
({\ref{eq:Reaction}).

Calculation of the matrix element is straightforward
and analogous to the case of synchrotron neutrino-pair
emission ($e \rightarrow e + \nu + \bar{\nu}$~; see, e.g.,
\cite{yak81}). The final result is
%
\begin{eqnarray}
| V_{\rm f\,i} |^2 &=& \frac{(2\pi)^2 G_{\rm F}^2}{L_x L_z V^2} \,
                       \delta \left( p'_x + k'_x - p_x - k_x \right)
\nonumber \\
              &\times& \delta \left( p'_z + k'_z - p_z - k_z \right)
                       C^{ij} S_i S^\ast_j ,
\label{eq:MatrixElement}
\end{eqnarray}
where $G_{\rm F}$ is the Fermi weak-coupling constant,
$i$ and $j$ are four-tensor indices ($i,j=0,1,2,3$),
\begin{equation}
   C^{ij} = \frac{1}{\omega \omega'}
            \left( k^i k'^j + k'^i k^j - g^{ij} k^l k'_l +
            {\rm i} e^{ijlm} k_l k'_m \right) ,
\label{eq:Cij}
\end{equation}
$k^i=(\omega,{\bf k})$, $e^{ijlm}$ is the antisymmetric
unit tensor, and $S_i$ is the transition amplitude of the
four-current for the NES, given
by Eq.~(\ref{eq:Amplitudes}). We introduce the cross section
$ {\rm d} \sigma_{\rm f\,i} = V {\rm d} w_{\rm f\,i} $
of the NES by
\widetext
%
\begin{equation}
{\rm d} \sigma_{\rm f\,i} =
\frac{G_{\rm F}^2}{(2\pi)^2}
\, C^{ij} S_i S^\ast_j \,
\delta \left( \varepsilon' + \omega' - \varepsilon - \omega \right)
\delta \left( p'_x + k'_x - p_x - k_x \right)
\delta \left( p'_z + k'_z - p_z - k_z \right)
\, {\rm d} p'_x \, {\rm d} p'_z \, {\rm d} {\bf k}',
\label{eq:PCS}
\end{equation}
and perform averaging over initial and summation over final
electron states,
%
\begin{equation}
{\rm d} \sigma = \frac{b}{(2\pi)^2}
                 \sum_{ n = 0 }^\infty
                 \int_{-\infty}^{+\infty} {\rm d} p_z
                 \sum_{ n' = 0 }^\infty
                 \int_{-\infty}^{+\infty} {\rm d} p'_x
                 \int_{-\infty}^{+\infty} {\rm d} p'_z
                 \sum_{ s, s' = \pm 1 }
                 f ( 1 - f' ) \, {\rm d} \sigma_{\rm f\,i},
\label{eq:DS}
\end{equation}
where
%
\begin{equation}
f = f(\varepsilon) =
\left[ \exp \left( \frac{\varepsilon-\mu}{T} \right) + 1 \right]^{-1}
\label{eq:FermiDirac}
\end{equation}
is the Fermi-Dirac distribution of the electrons in
the initial state,
while $f'=f(\varepsilon')$ is  same quantity for electrons
in the final state,
$T$ is temperature of matter
and $\mu$ is the electron chemical potential
(in units of $m_e c^2$). Note that the electron number density
(in units of $(m_e c / \hbar)^3$) is given by
%
\begin{equation}
n_e = \frac{b}{(2\pi)^2}
      \sum_{n=0}^\infty
      \int_{-\infty}^{+\infty} {\rm d} p_z
      \sum_{ s = \pm 1 } f
\label{eq:n_e}
\end{equation}
Taking into account that
${\rm d} {\bf k}' = \omega'^2 {\rm d} \omega' \, {\rm d} \Omega'$
(${\rm d} \Omega'$ being a solid angle element along ${\bf k}'$)
and introducing the quantities
\begin{equation}
    A_{ij} = \sum_{s,s' = \pm 1} S_i S_j^\ast , \; \;
    D = \varepsilon \varepsilon' C^{ij} A_{ij},
\label{eq:A_ij}
\end{equation}
we come to the following general expression for the differential
cross section of NES in dense magnetized plasma
\begin{eqnarray}
\frac{ {\rm d} \sigma }{ {\rm d} \omega' {\rm d} \Omega' }
&=& \frac{ G_{\rm F}^2 \, b \, \omega^{\prime 2} }{ (2 \pi)^4 }
    \sum_{n,n'=0}^\infty \int_{-\infty}^{+\infty}
    \frac{{\rm d} p_z}{\varepsilon \varepsilon'}
\nonumber \\
&\times& \delta \left( \varepsilon' + \omega'
                     - \varepsilon - \omega \right)
         f ( 1 - f' ) \, D .
\label{eq:CrossSection}
\end{eqnarray}
Explicit formulae for the quantities $A_{ij}$ and $D$
are given in Appendix~A. These quantities are expressed
in terms of the Laguerre functions $F_{n',n}(u)$
defined by Eq.~(\ref{eq:Fnn}). The Laguerre functions which
appear most frequently in the formulae
are denoted, for the sake of convenience,
as $F_1$, $F_2$, $F_3$, $F_4$:
they are defined in (\ref{eq:F1234}).

\section{Scattering cross section}
\label{Sect.2}
Equation (\ref{eq:CrossSection}) expresses the differential
scattering cross section as a sum over Landau
level numbers
of the initial and final electrons and an integral over
$p_z$. The integrand contains the energy conserving
delta function, so that  integration over $p_z$
can be done analytically.
For this purpose, we write
\begin{equation}
   \delta \left( \varepsilon' + \omega'
               - \varepsilon - \omega \right) =
   \sum_\lambda \frac{\varepsilon_\lambda \varepsilon'_\lambda}
                     {| \varepsilon'_\lambda p_{z \lambda}
			    - \varepsilon_\lambda p'_{z \lambda} |}
   \, \delta \left( p_z - p_{z \lambda} \right)
\label{eq:DeltaFunction}
\end{equation}
where $ \lambda $ enumerates all those
(``resonant") sets of energies and longitudinal
momenta $ \varepsilon, p_z, \varepsilon', p'_z $
which satisfy the conservation laws
%
\begin{eqnarray}
   \varepsilon' + \omega'  & = & \varepsilon + \omega ,
   \nonumber \\
   p'_z + k'_z & = & p_z + k_z .
   \label{eq:Conserva}
\end{eqnarray}

It is convenient to introduce the quantities
\begin{eqnarray}
   \Delta & = & \omega - \omega',
\nonumber \\
   \delta & = & k_z - k'_z = \omega \cos \vartheta
                           - \omega' \cos \vartheta',
\nonumber \\
   \xi    & = & \Delta^2 - \delta^2 =
   \omega^2 \sin^2 \vartheta + \omega^{\prime 2} \sin^2 \vartheta'
\nonumber \\
          & - & 2 \omega \omega'
	          ( 1 - \cos \vartheta \cos \vartheta' ),
\nonumber \\
   \chi   & = & - \xi,
\label{eq:Harmonics}
\end{eqnarray}
where $ \vartheta $ and $ \vartheta' $ are the polar angles of
{\bf k} and ${\bf k}'$, respectively. It is also suitable to
define the cyclotron harmonic number $ \nu = |n'-n| $
in such a way that $\nu \geq 0$.
An analysis of possible solutions, $\lambda$,
of Eqs.~(\ref{eq:Conserva}), to be used in
(\ref{eq:DeltaFunction}), reveals four different cases,
determined by signs of $ \Delta $ and $ \xi $.
We label these cases by A1, A2, B1, and B2, and discuss them below.

\widetext
{\bf Case A1} ($ \Delta \geq 0, \xi \geq 0 $).
In this case we always have $n > n'$. Eqs.~(\ref{eq:Conserva})
are then satisfied for $\nu \geq \nu_{\rm min}$ and
$n \leq n_{\rm max}$, where
\begin{equation}
\nu_{\rm min} = {\rm Int} \left[ \frac{ \sqrt{\xi} }{b}
     \left( \frac{ \sqrt{\xi} }{2} + 1 \right) \right] + 1 ,
\;\;\;\;
  n_{\rm max} = {\rm Int} \left\{ \frac{1}{2b} \left[
     \left( \frac{ \nu b }{ \sqrt{\xi} } -
     \frac{ \sqrt{\xi} }{2} \right)^2 - 1 \right] \right\} ,
\label{eq:Limits_A1}
\end{equation}
and where Int$(a)$ denotes the integral part of $a$.
For each $\nu$ and $n$ there are two solutions $\lambda=1,2$:
\begin{eqnarray}
   \varepsilon_1 &=&
   \frac{|\Delta|}{\xi} \left( \nu b - \frac{\xi}{2} \right) +
   \frac{|\delta|}{\xi}
   \sqrt{ \left( \nu b - \frac{\xi}{2} \right)^2
        - \varepsilon^2_{n0} \xi },
\nonumber \\
   p_{z1} &=& \left[
   \frac{|\delta|}{\xi} \left( \nu b - \frac{\xi}{2} \right) +
   \frac{|\Delta|}{\xi}
   \sqrt{ \left( \nu b - \frac{\xi}{2} \right)^2 - \varepsilon^2_{n0}
   \xi } \,
           \right] {\rm sgn} ( \Delta ) \, {\rm sgn} ( \delta ),
\nonumber \\
   \varepsilon_2 &=&
   \frac{ \displaystyle
   \delta^2 \varepsilon_{n0}^2
   + \left( \nu b - \frac{\xi}{2} \right)^2 }
   { \displaystyle
       |\Delta| \left( \nu b - \frac{ \xi }{ 2 } \right) +
       |\delta| \sqrt{ \left(
                \nu b - \frac{ \xi }{ 2 } \right)^2 -
		\varepsilon^2_{n0} \xi } },
\nonumber \\
   p_{z2} &=&
   \frac{ \displaystyle
       \Delta^2 \varepsilon_{n0}^2 - \left( \nu b - \frac{\xi}{2}
       \right)^2 }
     { \displaystyle
       |\delta| \left( \nu b - \frac{\xi}{2} \right) +
       |\Delta| \sqrt{ \left(
                \nu b - \frac{\xi}{2} \right)^2 -
		\varepsilon^2_{n0} \xi } } \,
    {\rm sgn} ( \Delta ) \, {\rm sgn} ( \delta ),
\nonumber \\
    \varepsilon'_{1,2} &=& \varepsilon_{1,2} + \Delta,
\nonumber \\
    p'_{z1,2} &=& p_{z1,2} + \delta.
\label{eq:EP_A1}
\end{eqnarray}
Final result reads
\begin{equation}
     \frac{ {\rm d} \sigma }{ {\rm d} \omega'
     {\rm d} \Omega' } =
     \frac{ G_{\rm F}^2 b }{ (2 \pi)^4 } \, \omega'^2
     \sum_{\nu = \nu_{\rm min}}^\infty
     \sum_{n=0}^{n_{\rm max}}
     \sum_{\lambda=1,2}
     \frac{ D \left( n, n+\nu,
                     \varepsilon_\lambda, \varepsilon'_\lambda,
                     p_{z \lambda}, p'_{z \lambda} \right) }
     { \sqrt{ \displaystyle
              \left( \nu b - \frac{\xi}{2} \right)^2 -
	      \varepsilon^2_{n0} \xi } }
     f \left( \varepsilon_\lambda \right) \left[ 1 - f
     \left( \varepsilon'_\lambda \right) \right] .
\label{eq:CrossSection_A1}
\end{equation}

{\bf Case A2} ($ \Delta \geq 0, \xi < 0 $).
In this case there are no restrictions on possible values
of $n$ and $n'$ and there is only a single solution
of Eq.~(\ref{eq:Conserva}). However, it is convenient
to separate this
solution into two sets corresponding to
$ n \leq n' $ and $ n > n' $ (labeled, respectively,
as ``1" and ``2"):
\begin{eqnarray}
   \varepsilon_1 &=&
   \frac{ \displaystyle
       \delta^2 \varepsilon_{n0}^2 + \left( \nu b + \frac{\chi}{2}
       \right)^2 }
      { \displaystyle
       |\Delta| \left( \nu b + \frac{\chi}{2} \right) +
       |\delta| \sqrt{ \left(
                \nu b + \frac{\chi}{2} \right)^2 +
		\varepsilon^2_{n0} \chi } },
\nonumber \\
   p_{z1} &=&
   \frac{ \displaystyle
       \Delta^2 \varepsilon_{n0}^2 - \left( \nu b + \frac{\chi}{2}
       \right)^2 }
     { \displaystyle
       |\delta| \left( \nu b + \frac{\chi}{2} \right) +
       |\Delta| \sqrt{ \left(
                \nu b + \frac{\chi}{2} \right)^2 + \varepsilon^2_{n0}
		\chi } } \,
   {\rm sgn} ( \Delta ) \, {\rm sgn} ( \delta ),
\nonumber \\
    \varepsilon'_1 &=& \varepsilon_1 + \Delta,
\nonumber \\
    p'_{z1} &=& p_{z1} + \delta,
\label{eq:EP_A21}
\end{eqnarray}
and
\begin{eqnarray}
   \varepsilon'_2 &=&
   \frac{|\Delta|}{\chi} \left( \nu b + \frac{\chi}{2}
   \right) +
   \frac{|\delta|}{\chi}
   \sqrt{ \left( \nu b + \frac{\chi}{2} \right)^2 +
   \varepsilon^2_{n'0} \chi },
\nonumber \\
   p'_{z2} &=& \left[
   \frac{|\delta|}{\chi} \left( \nu b + \frac{\chi}{2}
   \right) +
   \frac{|\Delta|}{\chi}
   \sqrt{ \left( \nu b + \frac{\chi}{2} \right)^2 +
   \varepsilon^2_{n'0} \chi } \,
       \right] {\rm sgn} ( \Delta ) \, {\rm sgn}
       ( \delta ),
\nonumber \\
   \varepsilon_2 &=& \varepsilon'_2 - \Delta,
\nonumber \\
   p_{z2} &=& p'_{z2} - \delta.
\label{eq:EP_A22}
\end{eqnarray}
The final expression for the cross section
can be written as
\begin{eqnarray}
\frac{ {\rm d} \sigma }{ {\rm d} \omega' {\rm d} \Omega' }
&=& \frac{ G_{\rm F}^2 b }{ (2 \pi)^4 } \, \omega'^2
    \left[
    \sum_{\nu=0}^\infty \sum_{n=0}^\infty
    \frac{ D \left( n, n+\nu,
                   \varepsilon_1, \varepsilon'_1,
                   p_{z 1}, p'_{z 1} \right) }
         { \displaystyle
           \sqrt{ \left( \nu b + \frac{\chi}{2} \right)^2 +
           \varepsilon^2_{n0} \chi } } \,
    f \left( \varepsilon_1 \right)
    \left[ 1 - f \left( \varepsilon'_1 \right)
    \right]
    \right.
\nonumber \\
&+& \left.
    \sum_{\nu=1}^\infty \sum_{n'=0}^\infty
    \frac{ D \left( n'+\nu, n',
                    \varepsilon_2, \varepsilon'_2, p_{z 2},
                    p'_{z 2} \right) }
         { \displaystyle
           \sqrt{ \left( \nu b + \frac{\chi}{2} \right)^2 +
           \varepsilon^2_{n'0} \chi } } \,
     f \left( \varepsilon_2 \right)
     \left[ 1 - f \left( \varepsilon'_2 \right)
     \right]
    \right].
\label{eq:CrossSection_A2}
\end{eqnarray}

{\bf Case B1} ($ \Delta < 0, \xi \geq 0 $).
In this case, only the electron transitions with $n > n'$
are allowed.
Analogously to the case A1, there exists the lowest cyclotron
harmonics, $\nu = \nu_{\rm min}$, and the highest
final Landau level $n' = n'_{\rm max}$,
\begin{equation}
\nu_{\rm min} = {\rm Int} \left[ \frac{ \sqrt{\xi} }{b}
     \left( \frac{ \sqrt{\xi} }{2} + 1 \right) \right] + 1 ,
\;\;\;\;
 n'_{\rm max} = {\rm Int} \left\{ \frac{1}{2b} \left[
     \left( \frac{ \nu b }{ \sqrt{\xi} } -
     \frac{ \sqrt{\xi} }{2} \right)^2 - 1 \right] \right\} .
\label{eq:Limits_B1}
\end{equation}
There are two roots of Eqs.~(\ref{eq:Conserva}) for each
$\nu$ and $n'$:
\begin{eqnarray}
     \varepsilon'_1 &=&
     \frac{|\Delta|}{\xi} \left( \nu b - \frac{\xi}{2} \right) +
     \frac{|\delta|}{\xi}
     \sqrt{ \left( \nu b - \frac{\xi}{2} \right)^2 -
     \varepsilon^2_{n'0} \xi },
\nonumber \\
     p'_{z1} &=& \left[
     \frac{|\delta|}{\xi} \left( \nu b - \frac{\xi}{2} \right) +
     \frac{|\Delta|}{\xi}
     \sqrt{ \left( \nu b - \frac{\xi}{2} \right)^2 -
     \varepsilon^2_{n'0} \xi } \,
     \right] {\rm sgn} ( \Delta ) \, {\rm sgn} ( \delta ),
\nonumber \\
     \varepsilon'_2 &=&
     \frac{ \displaystyle
     \delta^2 \varepsilon_{n'0}^2 + \left( \nu b - \frac{\xi}{2}
     \right)^2 }
     { \displaystyle
       |\Delta| \left( \nu b - \frac{\xi}{2} \right) +
       |\delta| \sqrt{ \left(
                \nu b - \frac{\xi}{2} \right)^2 - \varepsilon^2_{n'0}
		\xi } },
\nonumber \\
     p'_{z2} &=&
     \frac{ \displaystyle
       \Delta^2 \varepsilon_{n'0}^2 - \left( \nu b - \frac{\xi}{2}
       \right)^2 }
     { \displaystyle
       |\delta| \left( \nu b - \frac{\xi}{2} \right) +
       |\Delta| \sqrt{ \left(
                \nu b - \frac{\xi}{2} \right)^2 -
		\varepsilon^2_{n'0} \xi } } \,
     {\rm sgn} ( \Delta ) \, {\rm sgn} ( \delta ),
\nonumber \\
     \varepsilon_{1,2} &=& \varepsilon'_{1,2} - \Delta,
\nonumber \\
     p_{z1,2} &=& p'_{z1,2} - \delta.
\label{eq:EP_B1}
\end{eqnarray}
The cross section reads
\begin{equation}
   \frac{ {\rm d} \sigma }{ {\rm d} \omega'
   {\rm d} \Omega' } =
   \frac{ G^2 b }{ (2 \pi)^4 } \, \omega'^2
   \sum_{\nu = \nu_{\rm min}}^\infty
   \sum_{n'=0}^{n'_{\rm max}}
   \sum_{\lambda=1,2}
   \frac{ D \left( n'+\nu, n',
                   \varepsilon_\lambda, \varepsilon'_\lambda,
                   p_{z \lambda}, p'_{z \lambda} \right) }
        { \sqrt{ \displaystyle
         \left( \nu b - \frac{\xi}{2} \right)^2
	         - \varepsilon^2_{n'0} \xi }}
    f \left( \varepsilon_\lambda \right)
    \left[ 1 - f \left( \varepsilon'_\lambda \right) \right] .
\label{eq:CrossSection_B1}
\end{equation}

The square-root denominators in
Eqs.~(\ref{eq:CrossSection_A1}) and (\ref{eq:CrossSection_A2})
for the cases A1 ($\nu=n'-n$) and B1
($\nu=n-n'$), respectively, can be presented in the following
symmetric form
\begin{equation}
\sqrt{ \left( (n'-n) b - \frac{\xi}{2} \right)^2
      - \varepsilon^2_{n0}  \xi } =
\sqrt{ \left( (n-n') b - \frac{\xi}{2} \right)^2
      - \varepsilon^2_{n'0} \xi } =
\frac{1}{2}
\sqrt{ \left( \varepsilon_1^2 - \xi \right)
       \left( \varepsilon_2^2 - \xi \right) } ,
\label{eq:Symm_Form}
\end{equation}
where
$\varepsilon_{1,2} = | \varepsilon_{n'0} \mp \varepsilon_{n0} |$.
In these cases, the domain of summation over $\nu$
and the lowest Landau level number corresponds to the inequality
\begin{equation}
\xi \leq \xi_1 = \varepsilon_1^2 =
                 \frac{ 4 \nu^2 b^2 }{ (\varepsilon_{n'0} +
		 \varepsilon_{n0})^2 },
\label{eq:Domain}
\end{equation}
and the scattering cross section exhibits the square-root
singular features when $\xi$ approaches the thresholds $\xi_1$,
corresponding to the electron transitions $n \to n'=n+\nu$
(case A1)
and $n' \to n+\nu$ (case B1).
Such features are associated with the behaviour of
the density of states
for magnetized electron gas and are known to appear for, e.g.,
electromagnetic cyclotron radiation \cite{bez91}.

{\bf Case B2} ($ \Delta < 0, \xi < 0 $).
As in the case A2, there are no restrictions on possible
values of $n$ and $n'$, and single
solution of Eqs.~(\ref{eq:Conserva}) can be split into two sets
corresponding to $ n \leq n' $ and $ n > n' $
(we label them as ``1" and ``2", respectively):
\begin{eqnarray}
    \varepsilon_1 &=&
    \frac{|\Delta|}{\chi} \left( \nu b + \frac{\chi}{2}
    \right) +
    \frac{|\delta|}{\chi}
    \sqrt{ \left( \nu b + \frac{\chi}{2} \right)^2 +
    \varepsilon^2_{n0} \chi },
\nonumber \\
    p_{z1} &=& \left[
    \frac{|\delta|}{\chi} \left( \nu b + \frac{\chi}{2}
    \right) +
    \frac{|\Delta|}{\chi}
    \sqrt{ \left( \nu b + \frac{\chi}{2} \right)^2 +
    \varepsilon^2_{n0} \chi } \,
    \right] {\rm sgn} ( \Delta ) \, {\rm sgn} ( \delta ),
\nonumber \\
    \varepsilon'_1 &=& \varepsilon_1 + \Delta,
\nonumber \\
    p'_{z1} &=& p_{z1} + \delta,
\label{eq:EP_B21}
\end{eqnarray}
and
\begin{eqnarray}
    \varepsilon'_2 &=&
    \frac{ \displaystyle
    \delta^2 \varepsilon_{n'0}^2 + \left( \nu b + \frac{\chi}{2}
    \right)^2 }
    { \displaystyle
     |\Delta| \left( \nu b + \frac{\chi}{2} \right) +
     |\delta| \sqrt{ \left(
            \nu b + \frac{\chi}{2} \right)^2 + \varepsilon^2_{n'0}
	    \chi } },
\nonumber \\
    p'_{z2} &=&
    \frac{ \displaystyle
    \Delta^2 \varepsilon_{n'0}^2 - \left( \nu b + \frac{\chi}{2}
    \right)^2 }
    { \displaystyle
     |\delta| \left( \nu b + \frac{\chi}{2} \right) +
     |\Delta| \sqrt{ \left(
       \nu b + \frac{\chi}{2} \right)^2 + \varepsilon^2_{n'0}
       \chi } } \,
     {\rm sgn} ( \Delta ) \, {\rm sgn} ( \delta ),
\nonumber \\
     \varepsilon_2 &=& \varepsilon'_1 - \Delta,
\nonumber \\
     p_{z2} &=& p'_{z2} - \delta.
\label{eq:EP_B22}
\end{eqnarray}
The cross section can then be written as
\begin{eqnarray}
\frac{ {\rm d} \sigma }{ {\rm d} \omega' {\rm d} \Omega' }
&=& \frac{ G^2 b }{ (2 \pi)^4 } \, \omega'^2
    \left[
    \sum_{\nu=1}^\infty \sum_{n=0}^\infty
    \frac{ D \left( n, n+\nu,
                    \varepsilon_1, \varepsilon'_1,
			  p_{z 1}, p'_{z 1} \right) }
         { \displaystyle
           \sqrt{ \left( \nu b + \frac{\chi}{2} \right)^2 +
           \varepsilon^2_{n0} \chi } } \,
    f \left( \varepsilon_1 \right)
    \left[ 1 - f \left( \varepsilon'_1 \right) \right]
    \right.
\nonumber \\
&+& \left.
    \sum_{\nu=0}^\infty \sum_{n'=0}^\infty
    \frac{ D \left( n'+\nu, n',
                    \varepsilon_2, \varepsilon'_2,
			  p_{z 2}, p'_{z 2} \right) }
          { \displaystyle
            \sqrt{ \left( \nu b + \frac{\chi}{2} \right)^2 +
            \varepsilon^2_{n'0} \chi } } \,
    f \left( \varepsilon_2 \right)
    \left[ 1 - f \left( \varepsilon'_2 \right) \right]
    \right].
\label{eq:CrossSection_B2}
\end{eqnarray}

Similar to Eq.~(\ref{eq:Symm_Form}), the square-root
denominators in
Eqs.~(\ref{eq:CrossSection_A2}) and (\ref{eq:CrossSection_B2})
can also be written in the symmetric form
\begin{equation}
\sqrt{ \left( (n'-n) b + \frac{\chi}{2} \right)^2
- \varepsilon^2_{n0}  \chi } =
\sqrt{ \left( (n-n') b + \frac{\chi}{2} \right)^2
- \varepsilon^2_{n'0} \chi } =
\frac{1}{2}
\sqrt{ \left( \varepsilon_1^2 + \chi \right)
       \left( \varepsilon_2^2 + \chi \right) } .
\label{eq:Symm_Form1}
\end{equation}
%
Note that, in the case A2 (B2),
the contributions from the transitions with
$ n' \leq n $ ($ n \leq n' $)
decrease with decreasing  $ \chi $, and vanish at $ \chi = 0 $.
In each case, for $ \xi = \chi = 0 $, some
resonant electron states have infinite energies and therefore
do not contribute
to the cross section, but there are always the resonant states
with finite energies. Therefore,
for $ \xi = \chi = 0 $, expressions obtained in the cases
A1 and A2 (B1 and B2) give the same results
for the scattering cross section.

If $ \Delta = 0 $, then $ \xi $ can take only non-positive values,
and Eqs.~(\ref{eq:CrossSection_A2}) and (\ref{eq:CrossSection_B2})
derived for the cases A2 and B2, respectively, coincide.

\section{Zero field limit}
\label{Sect.3}
The neutrino-electron  scattering for $B=0$ has been considered
by numerous authors (see, e.g., \cite{mez93,yue76,tub75}, and
the references therein). For a degenerate
electron gas, the scattering cross section can be then presented
in a quite compact form involving standard Fermi integrals.
In view of the fact that we could not find in the literature a
complete and convincing derivation of the $B=0$ expression for
the scattering cross section, we preferred to rederive it in
Appendix~B. Of course, this expression for the scattering
cross section should be re-obtained as a limit of the expressions
presented in Section~\ref{Sect.2} in the case when the
quantization of the electron motion in the plane perpendicular
to {\bf B} becomes negligible. The effect of the magnetic field
is small, when the energy difference between neighboring Landau
levels is much smaller than the temperature of the matter,
$ \varepsilon_{n+1,0} - \varepsilon_{n,0} \ll T $.
Since
$ \varepsilon_{n+1,0} - \varepsilon_{n,0} =
2b / ( \varepsilon_{n+1,0} + \varepsilon_{n,0} ) $,
and the energies of the most populated electron states can be
approximated  by the electron chemical potential
(assuming $ \mu \gg 1 $),
$ \varepsilon_{n+1,0} \sim \varepsilon_{n,0} \sim \mu $,
we arrive at the condition $ b \ll \mu T $. Under such a condition,
the expressions given by Eqs.~(\ref{eq:CrossSection_A1}),
(\ref{eq:CrossSection_A2}), (\ref{eq:CrossSection_B1}), and
(\ref{eq:CrossSection_B2}) should converge to the $B=0$ expressions
given by Eqs.~(\ref{eq:NMSCS}) and (\ref{eq:Fin_Res_I_123}).

Analytical demonstration of such a convergence is quite
complicated, due to the strongly different forms of final
expressions for the scattering cross section in the case of a
magnetized plasma (Section~\ref{Sect.2}) and in the $B=0$ case
(Appendix~B), respectively. We will prove that our
general equation (\ref{eq:CrossSection}) reproduces correctly
the limit of $B \to 0$. To facilitate the proof, we transform the
expressions for $B=0$, introducing an {\it arbitrary}
$z$-axis and rearranging the order of integration over initial
and final momenta of the electron in the corresponding cylindrical
coordinates (see Appendix~C).

The zero field limit of Eq.~(\ref{eq:CrossSection}) can be
obtained using the quasi-classical expressions for the
Laguerre functions $F_{n'n}(u)$ which enter
the quantity $D$ (Appendix~A). In our limiting case,
high Landau levels
are involved, $n \gg 1$ and $n' \gg 1$,
and we can neglect the difference between $F_1^2$, $F_2^2$,
$F_3^2$ and
$F_4^2$. All these functions are then equal to (see, e.g.,
\cite{kam81})
%
\begin{equation}
F^2 = \frac{2 \, b}
           {\pi \, \sqrt{ (p_1^2 - q_\perp^2)
	   (q_\perp^2 - p_2^2)}} ,
\label{eq:QF^2}
\end{equation}
where $p_{1,2}=p'_\perp \pm p_\perp$. Then, using the recurrent
relations for the Laguerre functions
\cite{kam81}, we have
%
\begin{eqnarray}
p_\perp p'_\perp F_1 F_2 &=&
\frac{\left( p_\perp^2 - p_\perp^{\prime 2} \right)^2 -
      \left( p_\perp^2 + p_\perp^{\prime 2} \right)
      q_\perp^2}
     {2 q_\perp^2} \, F^2 ,
\nonumber \\
p_\perp p'_\perp F_3 F_4 &=&
\frac{p_\perp^2 + p_\perp^{\prime 2} - q_\perp^2}{2} \, F^2 ,
\nonumber \\
p_\perp \left( F_1 F_3 + F_2 F_4 \right) &=&
\frac{p_\perp^2 - p_\perp^{\prime 2} + q_\perp^2}{q_\perp}
\, F^2 ,
\nonumber \\
p'_\perp \left( F_1 F_4 + F_2 F_3 \right) &=&
\frac{p_\perp^2 - p_\perp^{\prime 2} - q_\perp^2}{q_\perp}
\, F^2 ,
\nonumber \\
p_\perp \left( F_1 F_3 - F_2 F_4 \right) &=&
p'_\perp \left( F_1 F_4 - F_2 F_3 \right) \;\; = \;\; 0 .
\label{eq:RR}
\end{eqnarray}
Substitution of Eqs.~(\ref{eq:QF^2}) and (\ref{eq:RR})
into Eq.~(\ref{eq:D}) gives the
quasi-classical expression
(\ref{eq:QCQ})
for the quantity $Q=D/(2 F^2)$.
Finally, replacing the summation over $n$ and $n'$ in
(\ref{eq:CrossSection}) by integration over
$p_\perp^2$ and $p_\perp^{\prime 2}$,
%
\begin{equation}
\sum_{n,n'} \rightarrow
\int \!\!\! \int
\frac{ {\rm d} p_\perp^2 {\rm d} p_\perp^{\prime 2} }
{ 4 \, b^2 }~,
\label{eq:Replacement}
\end{equation}
we get
Eq.~(\ref{eq:QCCS}), obtained for the case of zero magnetic field.
This strictly proves that our expression for
the cross section of
neutrino scattering on magnetized electrons,
Eq.~(\ref{eq:CrossSection}),
tends to the familiar cross section in the limit of
$B \to 0$.
%
\section{Numerical results}
\label{Sect.4}
We have calculated the
NES cross section using the expressions
presented in Section~\ref{Sect.2}, for a broad choice of physical
conditions in dense, hot, magnetized plasma. These conditions
were determined by the values of the dimensionless parameters
$\mu$, $T$, and $b$. The calculations were done for the scattering
of electron neutrino ($C_V \approx 0.96, C_A = 0.5$), although
the general expressions of Section~\ref{Sect.2}
can be applied for the scattering of
neutrino of any flavor.

We have used a simple approximate treatment of the singularities,
appearing in cases A1 and B1. Namely, when calculating
the square-root
denominators according to Eq.~(\ref{eq:Symm_Form}),
we were making the replacement
\begin{equation}
 \varepsilon_1^2 - \xi \to \left( \varepsilon_1 +
 \sqrt{\xi} \right)
 \sqrt{ \left( \varepsilon_1 - \sqrt{\xi} \right)^2 +
 \gamma^2 } .
 \label{eq:Resonant_Sqrt}
\end{equation}
Here, $\gamma$ has the meaning  of a width of the
electron energy levels,
determined by an appropriate mechanism of
de-excitation of the electron
states. For the interior of a neutron star, the most
important are the
electron-nucleus collisions. Assuming that the electrons
are degenerate
and neglecting the quantization of their transverse motion, we
may estimate the widths by the well-known result for the
frequency of the electron-nucleus Coulomb collisions \cite{yak80}
\begin{equation}
 \gamma_{\rm coll} \approx \frac{ 4 \alpha_{\rm f}^2 }
 { 3 \pi } Z \mu ,
 \label{eq:Width}
\end{equation}
where $\alpha_{\rm f} = 1/137$ is the fine structure constant,
$Z$ is the nucleus charge number, and $\mu$ is the
electron chemical
potential. In the range of temperatures and densities,
for which NES is most important during
the infall stage of gravitational collapse
($k_{\rm B}T \sim 1-2~{\rm MeV}$,
$\rho \sim 10^{10}-10^{12}~{\rm g~cm^{-3}}$, see
\cite{mez93}), we expect $Z \sim 30 - 35$.

In order to visualize the different domains of the NES in a
magnetized plasma, which were discussed in Section~\ref{Sect.2}
(cases A1, A2, B1, and B2), we present in
Fig.~\ref{F1} the differential cross section for
this process as a function of the initial neutrino energy.
The bottom panel of this figure
shows the dependence of the parameter $\xi$ on the energy
transfer. We also show the threshold values $\xi_1$
associated with the electron transition
from ($\omega > \omega'$, case A1) or to
($\omega < \omega'$, case B1)
the ground Landau level for different cyclotron harmonic
numbers $\nu$, and connect them with the corresponding
seesaw-like peaks,
associated with the square-root singularities
in the electron density of states.

It should also be noted, that peaks of
another kind appear in the cross section near the
boundaries separating
the domains A1 and A2, and B1 and B2, respectively.
They are associated
with those ``resonant" electron energies which grow infinitely at
$\xi$ approaching zero. For the cases A1 and B1, these energies
are labeled by ``2" and ``1", respectively
(see Eqs.~(\ref{eq:EP_A1}) and
(\ref{eq:EP_B1})). For the domain of $\xi < 0$,
the energies of the
electron transitions with $n'=n$ ($\nu=0$)
become infinite at $\xi \to 0$;
they are labeled by ``1" and ``2" for the cases A2 and B2,
respectively
(see Eqs~(\ref{eq:CrossSection_A2}) and
(\ref{eq:CrossSection_B2})).
Since with increasing energy the cross section
grows as energy squared,
the above discussed features produce narrow peaks
with amplitudes
determined by the condition $\varepsilon \approx \mu$
or $\varepsilon' \approx \mu$.
The peak shapes are also determined by the Fermi-Dirac
distribution of electrons: note that the
factor $f(1-f')$ is a peaked
function of initial (final) neutrino energy at
the low energy transfer ($|\Delta| \lesssim T$).
Two examples of such peaks are seen in Fig.~\ref{F2};
they are associated
there with the electron transitions
$n=2 \to n'=1$ (case B1), and $n=1 \to n'=1$
(case B2).

A sequence of plots in Fig.~\ref{F3} shows the
convergence of the scattering cross section
to the zero field limit.
Calculations were performed for different strengths of the
magnetic field, at fixed values
of the electron chemical potential and temperature which
are similar to those prevailing in the outer shells of
a collapsing core where NES is important.
The initial neutrino energies were also chosen to be
similar to
the mean neutrino energies in these
shells (see \cite{mez93}).
With the decrease of the magnetic field, the quantum
oscillations of the cross section become less pronounced,
and  deviations of its low and high initial neutrino
energy tails from the zero field limit
(see left panel of Fig.~\ref{F3}) disappear.

At $B=0$, the geometry of the NES can be described by
only one parameter,
which is the scattering angle $\theta$ between
the neutrino momenta, ${\bf k}'$ and {\bf k}.
At presence of the magnetic field, the geometry
is more complicated,
since $\theta$ is determined by three independent variables,
$\cos\theta = \cos\vartheta \cos\vartheta' +
\sin\vartheta \sin\vartheta' \cos(\alpha-\alpha')$.
The cross section becomes then quite sensitive
to a choice of $\vartheta$ and $\vartheta'$.
In particular, these angles determine the boundaries
of the domain A2$+$B2. Denoting them by $\omega_{\rm r,l}$
(at fixed $\omega'$)
and  by $\omega'_{\rm r,l}$ (at fixed $\omega$),
we can find from the
condition $\xi = 0$ the following relation,
\begin{eqnarray}
      \sin^2\vartheta  \, \frac{\omega_{\rm r,l}}{\omega'}
  &=& \sin^2\vartheta' \, \frac{\omega'_{\rm r,l}}{\omega} =
      1 - \cos\vartheta \cos\vartheta'
\nonumber \\
&\pm&
      \sqrt{ (1 - \cos(\vartheta-\vartheta'))
	       (1 + \cos(\vartheta+\vartheta')) } .
 \label{eq:Boundaries}
\end{eqnarray}
Note that the lower and the upper boundaries coincide at
$\vartheta=\vartheta'$
or $\vartheta=-\vartheta'$.

In Figs.~\ref{F4} and \ref{F5} we present the cross section
for the scattering
angle $\theta=90^\circ$, for two limiting cases, when the scattered
neutrino propagates in the plane ``{\bf k}--{\bf B}", or in the
direction perpendicular to this plane, respectively.
We have calculated, under these
kinematical conditions, the cross section as
the function of $\omega'$ at different polar angles $\vartheta$
of the incident neutrino.

In Fig.~\ref{F4}, eight examples of the cross section versus energy
plots are shown, which correspond to
$\theta=90^\circ$, and for $\vartheta$ varying from zero
to $315^\circ$ (with steps of $45^\circ$). As we see,
the shapes of all cross section plots are strikingly different.
The dependence of the cross section on $\vartheta$ in the case of
the scattering in the plane ``{\bf k}--{\bf B}" has the periodicity
of $360^\circ$, since at fixed scattering angle different
$\vartheta$ correspond to different $\vartheta'$, while the cross
section is determined by both polar angles.
Let us notice specific features of the plots
for $\vartheta=0$ and $\vartheta=180^\circ$, where
the domain A2$+$B2 in the cross section is the widest
($\omega'_{\rm l} = 0$, $\omega'_{\rm r} = 2 \omega$),
and the plots for $\vartheta=135^\circ$ and $\vartheta=315^\circ$,
when this domain vanishes completely.
For $\vartheta=135^\circ$ and $\vartheta=315^\circ$,
both $\Delta$ and $\xi$ are equal to zero
at the boundary of the domains A1 and B1,
so that all the ``resonant" electron energies become
infinite (see Eqs.~({\ref{eq:EP_A1})
and (\ref{eq:EP_B1})). This leads to the deep minima in the
cross section around the point $\omega' = \omega$, at which the
cross section is strictly equal to zero.

When the scattered neutrino propagates transversely
to {\bf B} and the angle of scattering, $\theta$, is fixed, the
shape of the cross section is determined only by the polar angle
of the incident neutrino. In this case, the cross section
is invariant with respect to the replacement of $\vartheta$ by
$360^\circ - \vartheta$, which is
demonstrated in Fig.~\ref{F5}. One can notice that the
influence of magnetic field
on the cross section is most significant at $\vartheta=90^\circ$,
where the quantum (oscillatory) features appear
near the maximum of the cross section,
and the domain without quantum singularities (A2$+$B2) vanishes.
Comparison of this example with those shown in Fig.~\ref{F4}
allows one to conclude,
that the quantum oscillations are most pronounced in the
cross section, when both incident and scattered neutrino
move transversely to the strong magnetic field.

To conclude our analysis, we show in Fig.~\ref{F6} how the presence
of the magnetic field breaks the invariance of cross the section
with respect to the exchange of directions of {\bf k} and ${\bf k}'$.
We have set $\alpha=\alpha'=45^\circ$
and computed the cross section for different pairs of
$\vartheta \leftrightarrow \vartheta'$.
One can see that the regions of prominent quantum oscillations are
different for all pairs $\vartheta \leftrightarrow \vartheta'$.
On the other hand, one can notice that the effect
of anisotropy is correlated with the specific direction of the
propagation of one of the neutrinos (incident or scattered)
relative  to the magnetic field. Namely,
in the case when the initial (final) neutrino momentum is
parallel to {\bf B}, there is a strong decrease of the cross
section for the neutrino energy loss (Fig.~\ref{F6}a). This effect
is reversed (increase of cross section for energy loss), if the
initial or final neutrino momentum is anti--parallel to {\bf B}
(Fig.~\ref{F6}b, see also examples shown in Fig.~\ref{F4}).

Numerical results, presented in this section, were obtained for
selected sets of the input parameters entering the expressions
for the scattering cross section. These results were used to
show some characteristic features of the neutrino-electron
scattering in a hot, dense magnetized plasma. Both the formulae
and numerical results are quite complex. Therefore, to make our
results usable for interested workers in this field, the Fortran
program is available upon request from one of the authors
(V.G.Bezchastnov, e-mail: vit@astro.ioffe.rssi.ru).

%
\section{Discussion and conclusions}
\label{Sect.5}
We have considered the NES in
a dense, hot, degenerate plasma, under conditions expected
to prevail during the
infall phase of the gravitational collapse, and assuming the
presence of a strong magnetic field.
Our results show that the magnetic field modifies the
energy dependence of the scattering cross section, and
introduces an
anisotropy correlated with the direction of {\bf B}. The
quantization of the electron motion induces the appearance of
an oscillatory behavior of the cross section as a function of the
energy transfer. The amplitude of the oscillations and the
distance between them strongly depend on the magnetic
field strength.

Among the neutrino processes which are relevant for the infall
phase of gravitational collapse, the neutrino-electron
scattering is
the most sensitive one to the presence of a strong magnetic
field. It should
be noted, however, that
a very strong magnetic field, exceeding $10^{15}~$G, influences
also the rate of electron captures on protons, which produce
electron neutrinos during the infall stage. Generally, the
presence of a strong magnetic field decreases the rate of electron
captures \cite{lai91}. However, this effect can be
neglected for the densities greater than $10^{10}~{\rm
g~cm^{-3}}$, and therefore will not play a significant role for
the collapse scenario, where the electron captures take place
mainly at the density close to the trapping density.

The presence of a strong magnetic field implies a very
strong sensitivity of the differential cross section for the
NES on the spatial direction of
neutrino momenta, as well as on the energy transfer. This could
lead to anisotropies and asymmetry in the heating of matter, due to
anisotropic energy transfer from neutrinos to matter, as well as
to the anisotropic and/or asymmetric momentum transfer to the
matter. Both effects could contribute to asymmetry
and anisotropy of the subsequent explosion of the outer
layers of collapsing core.

Both anisotropy and sensitivity to the energy transfer become
dramatically large  for a sufficiently strong magnetic
field, which by itself could influence the dynamics of
collapsing layers of the stellar core. Let us define the limiting
value of the magnetic field, $B_{\rm dyn}$, by $B_{\rm
dyn}^2/(8\pi)= P_{\rm matter}$.
For $B \ll B_{\rm dyn}$
the direct effect of the magnetic field on the dynamics of a layer
can be neglected, while for
$B \gtrsim B_{\rm dyn}$ the magnetic field modifies in an
essential way the equation of state of a given layer of the
collapsing core. We have $B_{\rm dyn}=5\times
10^{15}~(P_{\rm matter}/10^{30}~
{\rm dyn~cm^{-2}})^{1/2}~{\rm G}$.
Using the
results of numerical simulations of \cite{mez93},
we find that for the layers in which the neutrino-electron
scattering is important, $B_{\rm dyn}\sim 10^{15}~$G. Our
numerical results show, that the effects of the magnetic field on
the NES become very strong for a
field strength of the order of $B_{\rm dyn}$. It should be
mentioned, however, that the fact that the local magnetic field
strength exceeds $B_{\rm dyn}$ does not mean that the magnetic field
influences in a crucial way  the dynamics of collapse: the
effect of the local pressure on
the dynamics of collapsing outer layers is not very important
during the infall phase.

\acknowledgements{This work was supported in part by the
Polish State Committee for Scientific Research (KBN), grant
No.~P304 014 07, as well as by the INTAS grant No.~94-3834 and
Soros grant No.~R6A000.
One of the authors (V.G.B.) is grateful for excellent conditions and
hospitality during his stay at the
N.~Copernicus Astronomical Center.
We express our gratitude to D.G.~Yakovlev for
useful discussions and for helpful remarks concerning the analytical
part of this work.}

\appendix
\section{}
\label{Ap.A}
The structure of the transition amplitude of the four-current
$S^i = (S^0, {\bf S})$ for the NES is similar to that for the
neutrino synchrotron process (see, e.g., \cite{kam92,yak81}).
We have
\begin{eqnarray}
        S^{0} & = & [C_{V}(\alpha \alpha' +
	s s' \beta \beta' )
        -C_{A}(\alpha s' \beta' + \alpha'
	s \beta)] a^{0},
    \nonumber \\
        {\bf S} & = & [C_{V}(\alpha s' \beta' +
	\alpha' s \beta )
        -C_{A}(\alpha \alpha' + s s' \beta \beta')]
	{\bf a},
    \label{eq:Amplitudes}
\end{eqnarray}
where
\begin{eqnarray}
        a^{0} & = & AA'F_3+s s' B B' F_4,
    \nonumber \\
        a^{1} & = & A'B s e^{{\rm i} \varphi} F_1
        + s' B' A e^{-{\rm i} \varphi} F_2,
    \nonumber \\
        a^{2} & = & -iA'B s e^{{\rm i} \varphi} F_1
        +i s' B' A e^{-{\rm i} \varphi} F_2,
    \nonumber \\
        a^{3} & = & AA'F_3-s s' B B' F_4.
    \label{eq:SmallA}
\end{eqnarray}
In the above formulae,
\begin{eqnarray}
\pmatrix{\alpha\cr \beta\cr} &=& {1\over \sqrt{2}}
\Big(1\pm \frac{1}{\varepsilon} \Big)^{1/2},
\nonumber \\
         \pmatrix{A\cr B\cr} &=& {1\over\sqrt{2}}
\Big(1\pm \frac{s p_z}{\sqrt{\varepsilon^2-1}}\Big)^{1/2},
\end{eqnarray}
\begin{eqnarray}
    F_1 &=& F_{n'-1,n}(u), \; \; F_2 = F_{n',n-1}(u),
\nonumber \\
    F_3 &=& F_{n'-1,n-1}(u), \; \; F_4 = F_{n',n}(u),
\label{eq:F1234}
\end{eqnarray}
and
\begin{equation}
    F_{n',n}(u)=
    \left( {n'! \over n!} u^{n-n'} \right)^{1/2} {\rm e}^{-u/2}
    L_{n'}^{n-n'}(u)
\label{eq:Fnn}
\end{equation}
is a Laguerre function, $L_p^q(u)$
being an associated Laguerre polynomial.
In Eq.~(\ref{eq:Amplitudes}) $C_V$ and $C_A$ are
the vector and axial vector
weak interaction constants, respectively. For the
scattering of electron neutrino
(via charged and neutral currents), one has
$C_V = 2 \sin^2 \theta_W +0.5$ and $C_A= 0.5$,
while for the scattering of muonic or tauonic
neutrinos (neutral currents only),
$C'_V = 2 \sin^2 \theta_W - 0.5$ and $C'_A = -0.5$.
Here $\theta_W$ is the Weinberg angle,
$\sin^2 \theta_W \simeq 0.23$.

The only difference with respect to the case of the
neutrino synchrotron radiation,
is that in our case $ \varphi $ in Eq.~(\ref{eq:SmallA}) is
an angle between $ {\bf k}'_\perp - {\bf k}_\perp $ and
the $x$-axis, and the argument of the Laguerre functions is
$u=({\bf k}'_\perp - {\bf k}_\perp)^2/(2b)$.

Calculation of the 4-tensor $A_{ij}$ yields:
\widetext
\begin{eqnarray}
    A_{00} + A_{33}
    &=&
    \left(C_V^2 + C_A^2 \right)
    \left( 1 + \frac{p_z p'_z}{\varepsilon \varepsilon'} \right)
    \left( F_3^2 + F_4^2 \right) -
    2 C_V C_A
    \left( \frac{p_z}{\varepsilon} + \frac{p'_z}{\varepsilon'} \right)
    \left( F_3^2 - F_4^2 \right),
\nonumber \\
    A_{00} - A_{33}
    &=&
    2 \left( C_V^2 + C_A^2 \right)
    \frac{p_\perp p'_\perp}{\varepsilon \varepsilon'} F_3 F_4 +
    \frac{C_V^2-C_A^2}{\varepsilon \varepsilon'}
    \left( F_3^2 + F_4^2 \right),
\nonumber \\
    A_{11} + A_{22}
    &=&
    \left( C_V^2 + C_A^2 \right)
    \left( 1 - \frac{p_z p'_z}{\varepsilon \varepsilon'} \right)
    \left( F_1^2 + F_2^2 \right) -
    \frac{C_V^2-C_A^2}{\varepsilon \varepsilon'}
    \left( F_1^2 + F_2^2 \right) +
    2 C_V C_A
    \left( \frac{p_z}{\varepsilon} - \frac{p'_z}{\varepsilon'} \right)
    \left( F_1^2 - F_2^2 \right),
\nonumber \\
    A_{11} - A_{22}
    &=&
    2 \left( C_V^2 + C_A^2 \right)
    \frac{p_\perp p'_\perp}{\varepsilon \varepsilon'}
    \cos(2 \varphi) F_1 F_2 ,
\nonumber \\
    \frac{ A_{01} + A_{10} }{ \cos \varphi }
    &=& - \left( C_V^2 + C_A^2 \right)
    \left[ \frac{p_\perp}{\varepsilon}
           \left( F_1 F_3 + F_2 F_4 \right) +
           \frac{p'_\perp}{\varepsilon'}
           \left( F_1 F_4 + F_2 F_3 \right) \right]
\nonumber \\
    &+& 2 C_V C_A
    \left[ \frac{p_\perp p'_z}{\varepsilon \varepsilon'}
           \left( F_1 F_3 - F_2 F_4 \right) -
           \frac{p'_\perp p_z}{\varepsilon \varepsilon'}
           \left( F_1 F_4 - F_2 F_3 \right) \right] ,
\nonumber \\
    \frac{ A_{01} - A_{10} }{ {\rm i} \sin \varphi }
    &=& \left( C_V^2 + C_A^2 \right)
    \left[ \frac{p_\perp}{\varepsilon}
           \left( F_1 F_3 - F_2 F_4 \right) +
           \frac{p'_\perp}{\varepsilon'}
           \left( F_1 F_4 - F_2 F_3 \right) \right]
\nonumber \\
    &-& 2 C_V C_A
    \left[ \frac{p_\perp p'_z}{\varepsilon \varepsilon'}
           \left( F_1 F_3 + F_2 F_4 \right) -
           \frac{p'_\perp p_z}{\varepsilon \varepsilon'}
           \left( F_1 F_4 + F_2 F_3 \right) \right] ,
\nonumber \\
    \frac{ A_{02} + A_{20} }{ \sin \varphi }
    &=& - \left( C_V^2 + C_A^2 \right)
    \left[ \frac{p_\perp}{\varepsilon}
           \left( F_1 F_3 + F_2 F_4 \right) +
           \frac{p'_\perp}{\varepsilon'}
           \left( F_1 F_4 + F_2 F_3 \right) \right]
\nonumber \\
    &+& 2 C_V C_A
    \left[ \frac{p_\perp p'_z}{\varepsilon \varepsilon'}
           \left( F_1 F_3 - F_2 F_4 \right) -
           \frac{p'_\perp p_z}{\varepsilon \varepsilon'}
           \left( F_1 F_4 - F_2 F_3 \right) \right] ,
\nonumber \\
    \frac{ A_{02} - A_{20} }{ {\rm i} \cos \varphi }
    &=& - \left( C_V^2 + C_A^2 \right)
    \left[ \frac{p_\perp}{\varepsilon}
           \left( F_1 F_3 - F_2 F_4 \right) +
           \frac{p'_\perp}{\varepsilon'}
           \left( F_1 F_4 - F_2 F_3 \right) \right]
\nonumber \\
    &+& 2 C_V C_A
    \left[ \frac{p_\perp p'_z}{\varepsilon \varepsilon'}
           \left( F_1 F_3 + F_2 F_4 \right) -
           \frac{p'_\perp p_z}{\varepsilon \varepsilon'}
           \left( F_1 F_4 + F_2 F_3 \right) \right] ,
\nonumber \\
    A_{03} + A_{30}
    &=&
    - \left( C_V^2 + C_A^2 \right)
    \left( \frac{p_z}{\varepsilon} + \frac{p'_z}{\varepsilon'} \right)
    \left( F_3^2 + F_4^2 \right) +
    2 C_V C_A
    \left( 1 + \frac{p_z p'_z}{\varepsilon \varepsilon'} \right)
    \left( F_3^2 - F_4^2 \right),
\nonumber \\
    A_{03} - A_{30}
    &=&
    0,
\nonumber \\
    A_{12} + A_{21}
    &=&
    2 \sin(2 \varphi)
    \left( C_V^2 + C_A^2 \right)
    \frac{p_\perp p'_\perp}{\varepsilon \varepsilon'}
    F_1 F_2,
\nonumber \\
    \frac{ A_{12} - A_{21} }{ {\rm i} }
    &=&
    \left( C_V^2 + C_A^2 \right)
    \left( 1 - \frac{p_z p'_z}{\varepsilon \varepsilon'} \right)
    \left( F_1^2 - F_2^2 \right) -
    \frac{C_V^2 - C_A^2}{\varepsilon \varepsilon'}
    \left( F_1^2 - F_2^2 \right) +
    2 C_V C_A
    \left( \frac{p_z}{\varepsilon} - \frac{p'_z}{\varepsilon'} \right)
    \left( F_1^2 + F_2^2 \right) ,
\nonumber \\
    \frac{ A_{13} + A_{31} }{ \cos \varphi }
    &=& \left( C_V^2 + C_A^2 \right)
    \left[ \frac{p_\perp p'_z}{\varepsilon \varepsilon'}
           \left( F_1 F_3 + F_2 F_4 \right) +
           \frac{p'_\perp p_z}{\varepsilon \varepsilon'}
           \left( F_1 F_4 + F_2 F_3 \right) \right]
\nonumber \\
    &-& 2 C_V C_A
    \left[ \frac{p_\perp}{\varepsilon}
           \left( F_1 F_3 - F_2 F_4 \right) -
           \frac{p'_\perp}{\varepsilon'}
           \left( F_1 F_4 - F_2 F_3 \right) \right] ,
\nonumber \\
    \frac{ A_{13} - A_{31} }{ {\rm i} \sin \varphi }
    &=& \left( C_V^2 + C_A^2 \right)
    \left[ \frac{p_\perp p'_z}{\varepsilon \varepsilon'}
           \left( F_1 F_3 - F_2 F_4 \right) +
           \frac{p'_\perp p_z}{\varepsilon \varepsilon'}
           \left( F_1 F_4 - F_2 F_3 \right) \right]
\nonumber \\
    &-& 2 C_V C_A
    \left[ \frac{p_\perp}{\varepsilon}
           \left( F_1 F_3 + F_2 F_4 \right) -
           \frac{p'_\perp}{\varepsilon'}
           \left( F_1 F_4 + F_2 F_3 \right) \right] ,
\nonumber \\
    \frac{ A_{23} + A_{32} }{ \sin \varphi }
    &=& \left( C_V^2 + C_A^2 \right)
    \left[ \frac{p_\perp p'_z}{\varepsilon \varepsilon'}
           \left( F_1 F_3 + F_2 F_4 \right) +
           \frac{p'_\perp p_z}{\varepsilon \varepsilon'}
           \left( F_1 F_4 + F_2 F_3 \right) \right]
\nonumber \\
    &-& 2 C_V C_A
    \left[ \frac{p_\perp}{\varepsilon}
           \left( F_1 F_3 - F_2 F_4 \right) -
           \frac{p'_\perp}{\varepsilon'}
           \left( F_1 F_4 - F_2 F_3 \right) \right] ,
\nonumber \\
    \frac{ A_{23} - A_{32} }{ {\rm i} \cos \varphi }
    &=& - \left( C_V^2 + C_A^2 \right)
    \left[ \frac{p_\perp p'_z}{\varepsilon \varepsilon'}
           \left( F_1 F_3 - F_2 F_4 \right) +
           \frac{p'_\perp p_z}{\varepsilon \varepsilon'}
           \left( F_1 F_4 - F_2 F_3 \right) \right]
\nonumber \\
    &+& 2 C_V C_A
    \left[ \frac{p_\perp}{\varepsilon}
           \left( F_1 F_3 + F_2 F_4 \right) -
           \frac{p'_\perp}{\varepsilon'}
           \left( F_1 F_4 + F_2 F_3 \right) \right] ,
\label{eq:Aij}
\end{eqnarray}
where $ p_\perp = \sqrt{2nb} $ and $ p'_\perp =
\sqrt{2n'b} $.
The symmetric part of $A_{ij}$ corresponds to
Eqs.~(15) of \cite{kam92}.

Let $\vartheta$ and $\alpha$ be polar and azimuthal angles of
the incident neutrino momentum
($k_x= \omega \sin \vartheta \cos \alpha$,
$k_y = \omega \sin \vartheta \sin \alpha$,
$k_z= \omega \cos \vartheta$),
while $\vartheta'$ and $\alpha'$ be the same angles for the
scattered neutrino. Then
\begin{eqnarray}
    D
    &=& ( 1 + \cos \vartheta \cos \vartheta' )
    \left\{
    \left( C_V^2 + C_A^2 \right)
    \left( \varepsilon \varepsilon' + p_z p'_z \right) \left( F_3^2
    + F_4^2 \right) -
    2 C_V C_A \left( \varepsilon' p_z + \varepsilon p'_z \right)
    \left( F_3^2 - F_4^2 \right)
    \right\}                                      
\nonumber \\
    &+& ( 1 - \cos \vartheta \cos \vartheta' )
    \left\{
    \left( C_V^2 + C_A^2 \right)
    \left( \varepsilon \varepsilon' - p_z p'_z \right)
    \left( F_1^2 + F_2^2 \right) -
    \left( C_V^2 - C_A^2 \right)
    \left( F_1^2 +  F_2^2 \right)
    \right.
\nonumber \\
    &\;\;&
    \left.
    \;\;\;\;\;\;\;\;\;\;\;\;\;\;\;\;\;\;\;\;\;\;\;\;\;\;\;
    + 2 C_V C_A \left( \varepsilon' p_z - \varepsilon p'_z \right)
    \left( F_1^2 - F_2^2 \right)
    \right\}                                      
\nonumber \\
    &+& \sin \vartheta \sin \vartheta' \cos ( \alpha - \alpha' )
    \left\{
    2 \left( C_V^2 + C_A^2 \right) p_\perp p'_\perp F_3 F_4 +
    \left( C_V^2 - C_A^2 \right) \left( F_3^2 +  F_4^2 \right)
    \right\}                                      
\nonumber \\
    &+& 2 \sin \vartheta \sin \vartheta'
        \cos ( \alpha + \alpha' - 2 \varphi )
    \left( C_V^2 + C_A^2 \right) p_\perp p'_\perp F_1 F_2
\nonumber \\
    &-& ( \cos \vartheta + \cos \vartheta' )
    \left\{
    \left( C_V^2 + C_A^2 \right)
    \left( \varepsilon' p_z + \varepsilon p'_z \right)
    \left( F_3^2 + F_4^2 \right) -
    2 C_V C_A \left( \varepsilon \varepsilon' + p_z p'_z \right)
    \left( F_3^2 - F_4^2 \right)
    \right\}                                      
\nonumber \\
    &-& ( \cos \vartheta - \cos \vartheta' )
    \left\{
    \left( C_V^2 + C_A^2 \right)
    \left( \varepsilon \varepsilon' - p_z p'_z \right)
    \left( F_1^2 - F_2^2 \right) -
    \left( C_V^2 - C_A^2 \right) \left( F_1^2 - F_2^2 \right)
    +2 C_V C_A \left( \varepsilon' p_z - \varepsilon p'_z \right)
    \left( F_1^2 + F_2^2 \right) \right\}         
\nonumber \\
    &-& \left[ \sin \vartheta  \cos ( \alpha -  \varphi ) +
           \sin \vartheta' \cos ( \alpha' - \varphi )  \right]
    \left\{
    \left( C_V^2 + C_A^2 \right)
    \left[ p_\perp  \varepsilon' \left( F_1 F_3 + F_2 F_4 \right) +
    p'_\perp \varepsilon  \left( F_1 F_4 + F_2 F_3 \right) \right]
    \right.
\nonumber \\
    &\;\;& \left.
    \;\;\;\;\;\;\;\;\;\;\;\;\;\;\;\;\;\;\;\;\;\;\;\;\;\;\;\;\;\;
    \;\;\;\;\;\;\;\;\;\;\;\;\;\;\;\;\;\;\;\;\;\;\;\;\;\;\;\;\;\;
    - 2 C_V C_A
    \left[ p_\perp  p'_z \left( F_1 F_3 - F_2 F_4 \right) -
    p'_\perp p_z  \left( F_1 F_4 - F_2 F_3 \right) \right]
    \right\}                                      
\nonumber \\
    &+& \left[ \sin \vartheta  \cos ( \alpha  -  \varphi ) -
           \sin \vartheta' \cos ( \alpha'  - \varphi )  \right]
    \left\{
    \left( C_V^2 + C_A^2 \right)
    \left[ p_\perp  p'_z \left( F_1 F_3 - F_2 F_4 \right) +
    p'_\perp p_z  \left( F_1 F_4 - F_2 F_3 \right) \right]
           \right.
\nonumber \\
    &\;\;& \left.
    \;\;\;\;\;\;\;\;\;\;\;\;\;\;\;\;\;\;\;\;\;\;\;\;\;\;\;\;\;\;
    \;\;\;\;\;\;\;\;\;\;\;\;\;\;\;\;\;\;\;\;\;\;\;\;\;\;\;\;\;\;
    - 2 C_V C_A
    \left[ p_\perp  \varepsilon' \left( F_1 F_3 + F_2 F_4 \right) -
    p'_\perp \varepsilon  \left( F_1 F_4 + F_2 F_3 \right) \right]
    \right\}                                      
\nonumber \\
    &+& \left[ \cos \vartheta  \sin \vartheta'
               \cos ( \alpha' -  \varphi ) +
               \cos \vartheta' \sin \vartheta
		   \cos ( \alpha  - \varphi )  \right]
    \left\{
    \left( C_V^2 + C_A^2 \right)
    \left[ p_\perp  p'_z \left( F_1 F_3 + F_2 F_4 \right) +
     p'_\perp p_z  \left( F_1 F_4 + F_2 F_3 \right) \right]
           \right.
\nonumber \\
    &\;\;& \left.
    \;\;\;\;\;\;\;\;\;\;\;\;\;\;\;\;\;\;\;\;\;\;\;\;\;\;\;\;\;\;
    \;\;\;\;\;\;\;\;\;\;\;\;\;\;\;\;\;\;\;\;\;\;\;\;\;\;\;\;\;\;
    \;\;\;\;\;\;\;\;\;\;\;\;\;\;\;\;\;
    - 2 C_V C_A
    \left[ p_\perp  \varepsilon' \left( F_1 F_3 - F_2 F_4 \right) -
    p'_\perp \varepsilon  \left( F_1 F_4 - F_2 F_3 \right) \right]
       \right\}                                   
\nonumber \\
    &+& \left[ \cos \vartheta  \sin \vartheta'
               \cos ( \alpha' -  \varphi ) -
               \cos \vartheta' \sin \vartheta
		   \cos ( \alpha  - \varphi )  \right]
    \left\{
    \left( C_V^2 + C_A^2 \right)
    \left[ p_\perp  \varepsilon' \left( F_1 F_3 - F_2 F_4 \right) +
     p'_\perp \varepsilon  \left( F_1 F_4 - F_2 F_3 \right) \right]
           \right.
\nonumber \\
    &\;\;& \left.
    \;\;\;\;\;\;\;\;\;\;\;\;\;\;\;\;\;\;\;\;\;\;\;\;\;\;\;\;\;\;
    \;\;\;\;\;\;\;\;\;\;\;\;\;\;\;\;\;\;\;\;\;\;\;\;\;\;\;\;\;\;
    \;\;\;\;\;\;\;\;\;\;\;\;\;\;\;\;\;
    - 2 C_V C_A
    \left[ p_\perp  p'_z \left( F_1 F_3 + F_2 F_4 \right) -
    p'_\perp p_z  \left( F_1 F_4 + F_2 F_3 \right) \right]
    \right\}.                                      
\label{eq:D}
\end{eqnarray}
One can see that $D$ and, thus,
the scattering cross section depends on $\alpha - \alpha'$
rather than on the azimutal angles $\alpha$ and $\alpha'$
separately:
%
\begin{eqnarray}
   \sin ( \alpha - \varphi ) &=& \frac{k'_\perp}{q_\perp}
   \sin ( \alpha - \alpha' ) ,
\;\;\;\;
   \sin ( \alpha' - \varphi ) = \frac{k_\perp}{q_\perp}
   \sin ( \alpha - \alpha' ) ,
\nonumber \\
   \cos ( \alpha - \varphi ) &=& \frac{1}{q_\perp}
   \left[ k'_\perp \cos ( \alpha - \alpha' ) - k_\perp \right] ,
\;\;\;\;
   \cos ( \alpha' - \varphi ) = \frac{1}{q_\perp}
   \left[ k'_\perp - k_\perp \cos ( \alpha - \alpha' ) \right] ,
\nonumber \\
   \cos ( \alpha + \alpha' - 2 \varphi ) &=& \frac{1}{q_\perp^2}
   \left[ \left( k_\perp^2 + k_\perp^{\prime 2} \right)
          \cos ( \alpha - \alpha' ) - 2 k_\perp k'_\perp \right] ,
   \label{eq:Afa}
\end{eqnarray}
where
%
\begin{eqnarray}
   k_\perp &=& \omega \sin \vartheta ,
   \nonumber \\
   k'_\perp &=& \omega' \sin \vartheta' ,
   \nonumber \\
   q_\perp &=& \left[ k_\perp^2 + k_\perp^{\prime 2}
               - 2 k_\perp k'_\perp \cos ( \alpha - \alpha' )
	       \right]^{1/2} .
   \label{eq:Vtc}
\end{eqnarray}
Equations (\ref{eq:Afa}) cannot be used directly when
$ q_\perp = 0 $, i.e., when $ \omega = \omega' $ and
$ \alpha = \alpha' $ simultaneously.
If, however, $ \alpha =\alpha' $ one has
\begin{eqnarray}
   \sin ( \alpha - \varphi ) &=&
   \sin ( \alpha' - \varphi ) = 0,
   \nonumber \\
   \cos ( \alpha - \varphi ) &=&
   \cos ( \alpha' - \varphi ) = {\rm sgn} ( \omega' - \omega ) ,
   \nonumber \\
   \cos ( \alpha + \alpha' - 2 \varphi ) &=& 1,
   \label{eq:Afa1}
\end{eqnarray}
where $ {\rm sgn}(x) = 1 $ for $ x \geq 0 $
and $ {\rm sgn}(x) = -1 $ for $ x < 0 $.

When substituting $ D $ into Eq.~(\ref{eq:CrossSection}),
one should set
$ \varepsilon' = \varepsilon + \omega - \omega' $ and
$ p'_z = p_z + k_z - k'_z $.
%

\section{}
\label{Ap.B}
Let us consider the neutrino scattering on
non-magnetized electrons, whose
quantum states are determined by the
momentum ${\bf p}$ and helicity $s$.
The differential cross section
$ {\rm d} \sigma_{\rm f\,i} $ is calculated in the
straightforward manner, using the Fermi Golden rule.
After the averaging over the states of initial and
final electrons, we obtain
%
\begin{equation}
\frac{ {\rm d} \sigma }{ {\rm d} \omega' {\rm d} \Omega' } =
\frac{2 G_{\rm F}^2 \, \omega'}{(2 \pi)^5 \, \omega}
\left( C_1 I_1 + C_2 I_2 + C_3 I_3 \right),
\label{eq:NMSCS}
\end{equation}
where
$C_1=\left(C_V+C_A\right)^2$,
$C_2=\left(C_V-C_A\right)^2$,
$C_3=C_A^2-C_V^2$,
\widetext
%
\begin{eqnarray}
I_1 &=& I_1(k, k') =
        \int \frac{{\rm d} {\bf p} \, {\rm d} {\bf p}'}
	            {\varepsilon \varepsilon'} \,
        \delta^{(4)} \left( p' + k' - p - k \right) \,
        f (1-f') \, (pk)(p'k') ,
\;\;\;\;
I_2 = I_2(k, k') = I_1(-k', -k) ,
\nonumber \\
I_3 &=& I_3(k, k') =
        (kk') \int \frac{{\rm d} {\bf p} \, {\rm d} {\bf p}'}
	{\varepsilon \varepsilon'} \,
        \delta^{(4)} \left( p' + k' - p - k \right) \,
        f (1-f') .
\label{eq:I_123}
\end{eqnarray}
%
\noindent
Here, $p=(\varepsilon,-{\bf p})$ and $k=(\omega,-{\bf k})$
are the four-momenta of electron and neutrino before scattering, and
$p'=(\varepsilon',-{\bf p}')$ and $k=(\omega',-{\bf k}')$
are the 4-momenta of these particles after scattering,
$f=f(\varepsilon)$ and $f'=f(\varepsilon')$ are the Fermi-Dirac
functions given by Eq.~(\ref{eq:FermiDirac}), with
$\varepsilon=\sqrt{1+p^2}$ and $\varepsilon'=\sqrt{1+p'^2}$.

The four-dimensional delta-function in Eq.~(\ref{eq:I_123})
describes the conservation of momentum,
%
\begin{equation}
{\bf p} - {\bf p}'= {\bf q} ,
\label{eq:CTM}
\end{equation}
where ${\bf q} = {\bf k}' - {\bf k}$, and of energy,
%
\begin{equation}
\varepsilon' - \varepsilon = \Delta ,
\label{eq:CE}
\end{equation}
where $\Delta = \omega - \omega'$. Let us first perform
the integration over ${\bf p}'$. We use the
spherical coordinates with the polar axis along {\bf q}
to integrate over {\bf p}. This yields
\widetext
%
\begin{eqnarray}
I_1 &=& I_1( \omega,   \omega', \cos\theta) =
2 \pi \int_0^\infty \frac{p^2 {\rm d} p}{\varepsilon \varepsilon'} \,
      f (1-f')
      \int_{-1}^1 {\rm d} \eta \,
	\delta (\varepsilon'-\varepsilon-\Delta) \,
	I(\varepsilon, \varepsilon'),
\nonumber \\
I_2 &=& I_2( \omega,   \omega', \cos\theta) =
        I_1(-\omega', -\omega,  \cos\theta) ,
\nonumber \\
I_3 &=& I_3( \omega,   \omega', \cos\theta) =
2 \pi \omega \omega' (1-\cos\theta)
      \int_0^\infty \frac{p^2 {\rm d} p}
	                   {\varepsilon \varepsilon'} \, f (1-f')
      \int_{-1}^1 {\rm d} \eta \,
	\delta (\varepsilon'-\varepsilon-\Delta ) .
\label{eq:I_123CS}
\end{eqnarray}
%
\noindent
Here, $\theta$ is the angle between ${\bf k}$ and ${\bf k}'$,
$\eta$ is cosine of the polar angle of the electron momentum
${\bf p}$, and
%
\begin{equation}
I(\varepsilon, \varepsilon') = \frac{1}{2\pi} \int_0^{2\pi}
{\rm d} \beta \, (pk)(p'k') ,
\label{eq:I}
\end{equation}
where $\beta$  is the azimutal angle of ${\bf p}$.

Furthermore, we transform the energy conserving
delta-function in Eq.~(\ref{eq:I_123CS}) as:
%
\begin{equation}
\delta ( \varepsilon'-\varepsilon-\Delta ) =
  \frac{\varepsilon'}{q \sqrt{\varepsilon^2-1}} \,
  \delta ( \eta - \eta_0 ) ,
\label{eq:TDEF}
\end{equation}
where
%
\begin{equation}
\eta_0 = \frac{\varepsilon \Delta - \omega \omega' (1-\cos\theta)}
{q \sqrt{\varepsilon^2-1}} ,
\label{eq:ETA_0}
\end{equation}
$\varepsilon'=\varepsilon + \Delta$, and
%
\begin{equation}
q = \sqrt{\omega^2 + \omega^{\prime 2}
    - 2 \omega \omega' \cos\theta} .
\label{eq:Value_of_q}
\end{equation}
The condition $\eta_0^2 = 1$ determines the minimal electron
energy, which contributes to the integrals (\ref{eq:I_123CS}),
%
\begin{equation}
\varepsilon_{\rm min} = \frac{1}{2}
\left( q\sqrt{1+\frac{2}{\omega\omega'(1-\cos\theta)}}
      - \Delta \right) .
\label{eq:Minimal_Energy}
\end{equation}
Using (\ref{eq:TDEF}), we obtain
%
\begin{equation}
I_3 = 2 \pi \frac{\omega \omega'}{q} (1-\cos\theta)
      \int_{\varepsilon_{\rm min}}^\infty {\rm d} \varepsilon \,
	f (1-f') .
\label{eq:I_3}
\end{equation}

Evaluation of $I_1$ is more complicated because of the
presence of the integral
(\ref{eq:I}). The latter integral is equal to
%
\begin{equation}
I = \varepsilon \omega \varepsilon' \omega' -
    \varepsilon \omega p'_{{\scriptscriptstyle \parallel}}
                       k'_{{\scriptscriptstyle \parallel}} -
    \varepsilon' \omega' p_{{\scriptscriptstyle \parallel}}
                         k_{{\scriptscriptstyle \parallel}} +
    p_{{\scriptscriptstyle \parallel}}
    k_{{\scriptscriptstyle \parallel}}
    p'_{{\scriptscriptstyle \parallel}}
    k'_{{\scriptscriptstyle \parallel}} +
    \frac{1}{2} \, p_{\rm tr}^2 k_{\rm tr}^2 ,
\label{eq:Result_for_I}
\end{equation}
where the subscripts ``$\scriptstyle \parallel$" and ``tr"
denote the vector components parallel and perpendicular
to {\bf q}, respectively. The electron momentum components
determined by conservation laws are given by
%
\begin{eqnarray}
p_{{\scriptscriptstyle \parallel}} &=&
\frac{1}{2q} \, \left( \varepsilon^{\prime 2}
                      - \varepsilon^2 - q^2 \right) ,
\nonumber \\
p'_{{\scriptscriptstyle \parallel}} &=&
\frac{1}{2q} \, \left( \varepsilon^{\prime 2}
                      - \varepsilon^2 + q^2 \right) ,
\label{eq:EMC}
\\
p_{\rm tr}^2 &=&
\frac{1}{2q^2} \,
\left[ -q^4 + 2 (\varepsilon^2 + \varepsilon^{\prime 2} + 2)^2 q^2 -
       \left( \varepsilon^2 - \varepsilon^{\prime 2} \right)^2
\right] ,
\nonumber
\end{eqnarray}
and the neutrino momentum components expressed through
$\omega$, $\omega'$ and $\cos\theta$ are,
%
\begin{eqnarray}
k_{{\scriptscriptstyle \parallel}} &=&
\frac{\omega}{q} \, \left( \omega - \omega' \cos\theta \right) ,
\;\;\;
k'_{{\scriptscriptstyle \parallel}} \;\;=\;\;
\frac{\omega'}{q} \, \left( \omega \cos\theta - \omega' \right) ,
\nonumber \\
k_{\rm tr}^2 &=&
\frac{\omega^2 \omega^{\prime 2}}{q^2} \, \sin^2 \theta .
\label{eq:NMC}
\end{eqnarray}
Using Eqs.~(\ref{eq:EMC}) and (\ref{eq:NMC}), we obtain
%
\begin{equation}
I = 2 \pi \frac{\omega^2 \omega^{\prime 2}}{q^4}
    \left( 1 - \cos\theta \right)^2
    \left( a_0 + a_1 \varepsilon + a_2 \varepsilon^2 \right) ,
\label{eq:Final_Result_for_I}
\end{equation}
where
%
\begin{eqnarray}
a_0 &=& \omega^2 \left[
        \left( \omega - \omega' \cos\theta \right)^2 -
        \frac{\omega^{\prime 2} \sin^2 \theta}{2} -
        \frac{q^2 (1+\cos\theta)}{2 \omega^2 (1-\cos\theta)} \right] ,
\nonumber \\
a_1 &=& \omega \left[ 2 \omega^2 + \omega \omega' (3-\cos\theta) -
                   \omega^{\prime 2} (1+3\cos\theta) \right] ,
\nonumber \\
a_2 &=& \omega^2 + \omega^{\prime 2} + \omega \omega' (3+\cos\theta) .
\label{eq:a_012}
\end{eqnarray}
The coefficients $a_0$, $a_1$ and $a_2$ coincide,
respectively, with the
coefficients $C$, $B$ and $A$ given by Eqs.~(16)--(18)
of \cite{mez93}. Analogously to (\ref{eq:I_3}) we
obtain
%
\begin{eqnarray}
I_1 &=& 2 \pi \frac{\omega^2 \omega^{\prime 2}}{q^5} (1-\cos\theta)^2
\nonumber \\
&\times& \int_{\varepsilon_{\rm min}}^\infty {\rm d} \varepsilon \,
         \left( a_0 + a_1 \varepsilon + a_2 \varepsilon^2 \right) \,
	   f (1-f') .
\label{eq:I_1}
\end{eqnarray}

In order to integrate over electron energy in
Eqs.~(\ref{eq:I_3}) and (\ref{eq:I_1}), we use the energy
conservation, (\ref{eq:CE}), and the identity
%
\begin{equation}
f(1-f') = f_\gamma \left( - \frac{\Delta}{T} \right) (f-f') ,
\label{eq:Trans}
\end{equation}
where
%
\begin{equation}
f_\gamma (x) = \frac{1}{\exp(x)-1}
\label{eq:f_gamma}
\end{equation}
is independent of electron energy $\varepsilon$. Our final result
can be expressed in terms of the following integrals:
%
\begin{eqnarray}
\int_{\varepsilon_{\rm min}}^\infty {\rm d}
\varepsilon \, (f-f')  &=& T G_0 ,
\nonumber \\
\int_{\varepsilon_{\rm min}}^\infty {\rm d}
\varepsilon \, \varepsilon \,(f-f')  &=&
T^2 G_1 + T \varepsilon_{\rm min} G_0 ,
\label{eq:Integrals}
\\
\int_{\varepsilon_{\rm min}}^\infty {\rm d} \varepsilon \,
\varepsilon^2 \, (f-f')  &=&
T^3 G_2 + 2 T^2 \varepsilon_{\rm min} G_1
        + T \varepsilon_{\rm min}^2 G_0 ,
\nonumber
\end{eqnarray}
where
%
\begin{eqnarray}
G_k      &=& G_k(y,y') \;\; = \;\; F_k(y') - F_k(y) ,
\nonumber \\
y        &=&       y_0 + \frac{\Delta}{2T}, \;\;\;
y' \;\;\; = \;\;\; y_0 - \frac{\Delta}{2T},
\nonumber \\
y_0      &=& \frac{1}{2T}
\left( 2\mu - q\sqrt{1+\frac{2}{\omega\omega'(1-\cos\theta)}}
\, \right) ,
\label{eq:GF}
\end{eqnarray}
and
%
\begin{equation}
F_k(y) = \int_0^\infty \frac{\xi^k {\rm d} \xi}{1+\exp(\xi-y)}
\label{eq:Fermi_Integral}
\end{equation}
is a  standard Fermi integral.

Finally, using Eqs.~(\ref{eq:Trans}) and (\ref{eq:Integrals}),
we obtain
%
\begin{eqnarray}
I_1 &=& 2 \pi \frac{\omega^2 \omega^{\prime 2}}{q^5} (1-\cos\theta)^2
        f_\gamma \left( - \frac{\Delta}{T} \right) T
\nonumber \\
&\times& \left[ a_0 G_0 +
                a_1 \left( T G_1 + \varepsilon_{\rm min} G_0 \right)
         \right.
\nonumber \\
&\;\;& + \left. a_2 \left( T^2 G_2 + 2 T \varepsilon_{\rm min} G_1 +
                           \varepsilon_{\rm min}^2 G_0 \right) \right] ,
\nonumber \\
I_2 &=& I_2( \omega,   \omega', \cos\theta) =
        I_1(-\omega', -\omega,  \cos\theta) ,
\nonumber \\
I_3 &=& 2 \pi \frac{\omega \omega'}{q} (1-\cos\theta)
        f_\gamma \left( - \frac{\Delta}{T} \right) T G_0 .
\label{eq:Fin_Res_I_123}
\end{eqnarray}
This result coincides with Eq.~(13) of \cite{mez93}.
%

\section{}
\label{Ap.C}
In this appendix, we transform the general expression for the NES
cross section at $B=0$, introducing an arbitrary $z$-axis and
integrating over orientations of the electron momenta in the
corresponding perpendicular plane.
Let $\beta$ and $\beta'$ be the azimutal angles of the vectors
${\bf p}_\perp$ and ${\bf p}'_\perp$, respectively,
and $\chi$ be the angle between
${\bf p}_\perp$ and ${\bf p}'_\perp$. It is evident that
%
\begin{equation}
{\rm d} {\bf p} \, {\rm d} {\bf p}' = \frac{1}{4} \,
{\rm d} p_z \, {\rm d} p'_z  \,
{\rm d} p_\perp^2 \, {\rm d} p_\perp^{\prime 2} \,
{\rm d} \chi \, {\rm d} \beta .
\label{eq:TD}
\end{equation}
>From Eqs.~(\ref{eq:NMSCS}) and (\ref{eq:I_123}) we obtain
\widetext
%
\begin{equation}
\frac{ {\rm d} \sigma }{ {\rm d} \omega' {\rm d} \Omega' } =
\frac{ G_{\rm F}^2 \, \omega'}{2 (2 \pi)^5 \, \omega}
\int \!\!\! \int_0^\infty
{\rm d} p_\perp^2 {\rm d} p_\perp^{\prime 2}
\int_{-\infty}^{+\infty} \frac{{\rm d} p_z}{\varepsilon \varepsilon'}
\,
\delta \left( \varepsilon' + \omega' - \varepsilon - \omega \right)
f ( 1 - f' ) R ,
\label{eq:CrossSection1}
\end{equation}
where
%
\begin{eqnarray}
R &=& R_1 + R_2 + R_3 ,
\nonumber \\
R_1 &=& C_1
\int_0^{2 \pi} {\rm d} \chi
\int_0^{2 \pi} {\rm d} \beta \,
\delta \left( {\bf p}'_\perp + {\bf k}'_\perp -
              {\bf p}_\perp  - {\bf k}_\perp \right)
(pk)(p'k'),
\nonumber \\
R_2 &=& C_2
\int_0^{2 \pi} {\rm d} \chi
\int_0^{2 \pi} {\rm d} \beta \,
\delta \left( {\bf p}'_\perp + {\bf k}'_\perp -
              {\bf p}_\perp  - {\bf k}_\perp \right)
(pk')(p'k),
\nonumber \\
R_3 &=& C_3 \, (kk')
\int_0^{2 \pi} {\rm d} \chi
\int_0^{2 \pi} {\rm d} \beta \,
\delta \left( {\bf p}'_\perp + {\bf k}'_\perp -
              {\bf p}_\perp  - {\bf k}_\perp \right) .
\label{eq:R_123}
\end{eqnarray}
%

Our aim is to integrate over $\chi$ and $\beta$ in
Eq.~(\ref{eq:R_123}). We use the conservation of the
transverse electron momentum, which  implies
%
\begin{equation}
{\bf p}_\perp - {\bf p}'_\perp = {\bf q}_\perp ,
\label{eq:CTEM}
\end{equation}
where ${\bf q}_\perp = {\bf k}'_\perp - {\bf k}_\perp$. Combined
with the relation $\beta' = \beta + \chi$,
Eq.~(\ref{eq:CTEM}) leads to the following
equations for $\cos \beta$ and $\sin \beta$:
%
\begin{eqnarray}
(p_\perp - p'_\perp \cos \chi) \cos \beta +
p'_\perp \sin \chi \sin \beta &=& q_\perp \cos \varphi ,
\nonumber \\
- p'_\perp \sin \chi \cos \beta +
(p_\perp - p'_\perp \cos \chi) \sin \beta &=& q_\perp \sin \varphi ,
\label{eq:System}
\end{eqnarray}
where $\varphi$ is the azimutal angle of ${\bf q}_\perp$.
>From (\ref{eq:System}) we obtain
%
\begin{eqnarray}
\cos \beta &=& \frac{1}{q_\perp}
               \left[ (p_\perp - p'_\perp \cos \chi) \cos \varphi -
                      p'_\perp \sin \chi \sin \varphi \right] ,
\nonumber \\
\sin \beta &=& \frac{1}{q_\perp}
               \left[ p'_\perp \sin \chi \cos \varphi +
                      (p_\perp - p'_\perp \cos \chi)
		      \sin \varphi \right] ,
\label{eq:CSB}
\end{eqnarray}
and
%
\begin{eqnarray}
\cos \beta' &=& \frac{1}{q_\perp}
                \left[ (p_\perp \cos \chi - p'_\perp) \cos \varphi -
                       p_\perp \sin \chi \sin \varphi \right] ,
\nonumber \\
\sin \beta' &=& \frac{1}{q_\perp}
                \left[ p_\perp \sin \chi \cos \varphi +
                       (p_\perp \cos \chi - p'_\perp)
		       \sin \varphi \right] .
\label{eq:CSB'}
\end{eqnarray}
>From squared Eq.~(\ref{eq:CTEM}), we determine two possible
``resonant" values of
$\chi$ which contribute to the integrals over the directions of
${\bf p}_\perp$ and ${\bf p}'_\perp$:
%
\begin{eqnarray}
\cos \chi_\lambda &=&
\frac{p_\perp^2 + p_\perp^{\prime 2} - q_\perp^2}
{2 p_\perp p'_\perp} ,
\nonumber \\
\sin \chi_\lambda &=&
(-1)^\lambda \frac{\sqrt{ (p_1^2 - q_\perp^2)
(q_\perp^2 - p_2^2)}}
                  {2 p_\perp p'_\perp} ,
\label{eq:CSChi}
\end{eqnarray}
where $\lambda=1,2$ and $p_{1,2}=p'_\perp \pm p_\perp$. We can
rewrite the delta-function in Eqs.~(\ref{eq:R_123}) as
%
\begin{eqnarray}
\delta \left( {\bf p}'_\perp + {\bf k}'_\perp -
              {\bf p}_\perp - {\bf k}_\perp \right)
     &=& \frac{2}{\sqrt{ (p_1^2 - q_\perp^2) (q_\perp^2 - p_2^2)}}
\label{eq:TDF} \\
&\times& \sum_{\lambda=1,2}
         \delta ( \chi - \chi_\lambda ) \,
         \delta ( \beta - \beta_\lambda ),
\nonumber
\end{eqnarray}
where the angles $\beta_\lambda$ are determined by
Eqs.~(\ref{eq:CSB}) and (\ref{eq:CSB'}) with $\chi=\chi_\lambda$.
This yields
%
\begin{equation}
R_3 = \frac{4 \, C_3 \, \omega \, \omega' \, Q_3}
           {\sqrt{ (p_1^2 - q_\perp^2) (q_\perp^2 - p_2^2)}} ,
\label{eq:R_3}
\end{equation}
where
%
\begin{equation}
Q_3 =  1 - \cos \vartheta \cos \vartheta' -
\sin \vartheta \sin \vartheta' \cos(\alpha-\alpha') .
\label{eq:Q_3}
\end{equation}
In this case $\vartheta$ and $\vartheta'$ are the polar angles, and
$\alpha$ and $\alpha'$ are the azimuthal angles of
${\bf k}$ and ${\bf k}'$, respectively.

The expression for $R_1$ contains the quantity
\widetext
%
\begin{equation}
(pk)(p'k') =
\left( \varepsilon \omega - p_z k_z
      - {\bf p}_\perp {\bf k}_\perp \right)
\left( \varepsilon' \omega' - p'_z k'_z
      - {\bf p}'_\perp {\bf k}'_\perp \right) ,
\label{eq:(pk)(p'k')}
\end{equation}
in which ${\bf k}_\perp {\bf p}_\perp$ and
${\bf k}'_\perp {\bf p}'_\perp$
depend on the integration variable $\chi$. Using
Eqs.~(\ref{eq:CSB}) and (\ref{eq:CSB'}), we can explicitly
express these terms through $\chi$,
%
\begin{eqnarray}
{\bf p}_\perp {\bf k}_\perp &=&
\frac{p_\perp k_\perp}{q_\perp}     \left[
(p_\perp - p'_\perp \cos \chi) \cos (\alpha - \varphi) +
p'_\perp \sin \chi \sin (\alpha - \varphi) \right],
\nonumber \\
{\bf p}'_\perp {\bf k}'_\perp &=&
\frac{p'_\perp k'_\perp}{q_\perp}   \left[
(p_\perp \cos \chi - p'_\perp) \cos (\alpha' - \varphi) +
p_\perp \sin \chi \sin (\alpha' - \varphi) \right] .
\label{eq:SP}
\end{eqnarray}
This gives
%
\begin{equation}
R_1 = \frac{4 \, C_1 \, \omega \, \omega' \, Q_1}
           {\sqrt{ (p_1^2 - q_\perp^2) (q_\perp^2 - p_2^2)}} ,
\label{eq:R_1}
\end{equation}
where
%
\begin{eqnarray}
\omega \omega' Q_1
&=& ( \varepsilon \omega - p_z k_z )
    ( \varepsilon' \omega' - p'_z k'_z )
\label{eq:IQ_1} \\
&-& \frac{p'_\perp k'_\perp}{q_\perp}
    ( \varepsilon \omega - p_z k_z )
    (p_\perp \cos \chi - p'_\perp) \cos (\alpha' - \varphi)
  - \frac{p_\perp k_\perp}{q_\perp}
    ( \varepsilon' \omega' - p'_z k'_z )
    (p_\perp - p'_\perp \cos \chi) \cos (\alpha - \varphi)
\nonumber \\
&+& \frac{p_\perp p'_\perp k_\perp k'_\perp}{q^2_\perp}
    \left[ (p_\perp - p'_\perp \cos \chi)
           (p_\perp \cos \chi - p'_\perp)
           \cos (\alpha - \varphi) \cos (\alpha' - \varphi)
           + p_\perp p'_\perp \sin^2 \chi
            \sin (\alpha - \varphi) \sin (\alpha' - \varphi)
    \right] ;
\nonumber
\end{eqnarray}
$\cos\chi$ and $\sin\chi$ are given by
Eq.~(\ref{eq:CSChi}). Notice, that the terms
proportional to $\sin\chi$
do not contribute to $Q_1$ since they cancel out after summation
over $\lambda = 1,2$. Introducing
$\cos(\alpha-\alpha')$ and $\cos(\alpha+\alpha'-2\varphi)$
and using Eq.~(\ref{eq:CSChi}), we obtain
%
\begin{eqnarray}
Q_1
&=& \varepsilon \varepsilon'
    + p_z p'_z \cos \vartheta \cos \vartheta' -
    \varepsilon p'_z \cos \vartheta'
    - \varepsilon' p_z \cos \vartheta
\nonumber \\
&-& ( \varepsilon - p_z \cos \vartheta )
    \sin \vartheta' \cos (\alpha' - \varphi) \,
    \frac{p^2_\perp - p^{\prime 2}_\perp - q^2_\perp}{2 q_\perp}
    - ( \varepsilon' - p'_z \cos \vartheta' )
    \sin \vartheta \cos (\alpha - \varphi) \,
    \frac{p^2_\perp - p^{\prime 2}_\perp + q^2_\perp}{2 q_\perp}
\nonumber \\
&+& \sin \vartheta \sin \vartheta'
    \cos(\alpha + \alpha' - 2\varphi) \,
    \frac{ \left( p^2_\perp - p^{\prime 2}_\perp \right)^2 -
           \left( p^2_\perp + p^{\prime 2}_\perp \right) q^2_\perp }
         {4 q^2_\perp }
    + \sin \vartheta \sin \vartheta' \cos(\alpha - \alpha') \,
    \frac{p^2_\perp + p^{\prime 2}_\perp - q^2_\perp}{4} .
\label{eq:Q_1}
\end{eqnarray}
In analogy with (\ref{eq:R_1}), we have
%
\begin{equation}
R_2 = \frac{4 \, C_1 \, \omega \, \omega' \, Q_2}
           {\sqrt{ (p_1^2 - q_\perp^2) (q_\perp^2 - p_2^2)}} ,
\label{eq:R_2}
\end{equation}
where $Q_2$ is obtained from $Q_1$ by replacing
$\vartheta' \leftrightarrow \vartheta$,
$\alpha' \leftrightarrow \alpha$.
The final expression for the NES cross section reads
%
\begin{equation}
\frac{ {\rm d} \sigma }{ {\rm d} \omega' {\rm d} \Omega' } =
\frac{ 2 G_{\rm F}^2 \, \omega^{\prime 2} }{ (2 \pi)^5 }
\int \!\! \int
\frac{ {\rm d} p_\perp^2 {\rm d} p_\perp^{\prime 2} }
     { \sqrt{ (p_1^2 - q_\perp^2) (q_\perp^2 - p_2^2)} }
\int_{-\infty}^{+\infty} \frac{{\rm d} p_z}
                              {\varepsilon \varepsilon'} \,
\delta \left( \varepsilon' + \omega' - \varepsilon - \omega \right)
f ( 1 - f' ) Q ,
\label{eq:QCCS}
\end{equation}
where the integration domain over the transverse electron momenta
is restricted by the inequalities $p_2 \geq q_\perp$ and
$|p_1| \leq q_\perp$. The quantity
$Q = C_1 Q_1 + C_2 Q_2 + C_3 Q_3$ is explicitly
given by the following expression:
%
%
\begin{eqnarray}
Q &=& ( 1 + \cos \vartheta \cos \vartheta' )
      \left( C_V^2 + C_A^2 \right)
      \left( \varepsilon \varepsilon' + p_z p'_z \right)   
   +  ( 1 - \cos \vartheta \cos \vartheta' )
      \left[ \left( C_V^2 + C_A^2 \right)
             \left( \varepsilon \varepsilon' - p_z p'_z \right) -
             \left( C_V^2 - C_A^2 \right) \right]    
\nonumber \\
  &+& \sin \vartheta \sin \vartheta' \cos ( \alpha - \alpha' )
      \left[ \left( C_V^2 + C_A^2 \right)
             \frac{p_\perp^2 + p_\perp^{\prime 2}
             - q_\perp^2}{2} +
             \left( C_V^2 - C_A^2 \right) \right]           
\nonumber \\
  &+& \sin \vartheta \sin \vartheta'
      \cos ( \alpha + \alpha' - 2 \varphi )
      \left( C_V^2 + C_A^2 \right)
      \frac{\left( p_\perp^2 - p_\perp^{\prime 2}
            \right)^2 -
            \left( p_\perp^2 + p_\perp^{\prime 2} \right)
            q_\perp^2}
            {2 q_\perp^2}                                  
\nonumber \\
  &-& ( \cos \vartheta + \cos \vartheta' )
      \left( C_V^2 + C_A^2 \right)
      \left( \varepsilon' p_z + \varepsilon p'_z \right)   
   -  ( \cos \vartheta - \cos \vartheta' )
      2 C_V C_A \left( \varepsilon' p_z - \varepsilon p'_z \right)
\nonumber \\
  &-& \left[ \sin \vartheta  \cos ( \alpha -  \varphi ) +
             \sin \vartheta' \cos ( \alpha' - \varphi )  \right]
      \left( C_V^2 + C_A^2 \right)
      \left[ \varepsilon' \, \frac{p_\perp^2 -
                          p_\perp^{\prime 2} + q_\perp^2}
                         {2 q_\perp} +
             \varepsilon  \, \frac{p_\perp^2 - p_\perp^{\prime 2}
                          - q_\perp^2}
                         {2 q_\perp} \right]               
\nonumber \\
  &-& \left[ \sin \vartheta  \cos ( \alpha  -  \varphi ) -
             \sin \vartheta' \cos ( \alpha'  - \varphi )  \right]
      2 C_V C_A
      \left[ \varepsilon' \, \frac{p_\perp^2 - p_\perp^{\prime 2}
                          + q_\perp^2}
                         {2 q_\perp} -
             \varepsilon  \, \frac{p_\perp^2 - p_\perp^{\prime 2}
                          - q_\perp^2}
                         {2 q_\perp} \right]               
\nonumber \\
  &+& \left[ \cos \vartheta  \sin \vartheta'
             \cos ( \alpha' -  \varphi ) +
             \cos \vartheta' \sin \vartheta
		 \cos ( \alpha  - \varphi ) \right]
      \left( C_V^2 + C_A^2 \right)
      \left[ p'_z \, \frac{p_\perp^2 - p_\perp^{\prime 2}
                           + q_\perp^2}
                          {2 q_\perp} +
             p_z  \, \frac{p_\perp^2 - p_\perp^{\prime 2}
                           - q_\perp^2}
                          {2 q_\perp} \right]              
\nonumber \\
  &-& \left[ \cos \vartheta  \sin \vartheta'
             \cos ( \alpha' -  \varphi ) -
             \cos \vartheta' \sin \vartheta
		 \cos ( \alpha  - \varphi )  \right]
      2 C_V C_A
      \left[ p'_z \, \frac{p_\perp^2 - p_\perp^{\prime 2}
                           + q_\perp^2}
                          {2 q_\perp} -
             p_z  \, \frac{p_\perp^2 - p_\perp^{\prime 2}
                           - q_\perp^2}
                          {2 q_\perp} \right] .           
\label{eq:QCQ}
\end{eqnarray}
%

\vskip 0.5cm
\begin{figure}
\caption[]{The differential neutrino-electron
scattering cross section (in units of
$\sigma_0 = 4 G_{\rm F}^2 / \pi = 1.764 \times 10^{-44}$ cm$^2$)
as the function of the energy of the incident neutrino (top panel)
and the corresponding dependence of the parameter $\xi$
(bottom panel). The horizontal lines show the thresholds $\xi_1$
associated with the electron transition from (to) the ground Landau
level for different harmonics numbers $\nu$, and dashed vertical
lines label the square-root features of the cross section
corresponding to the thresholds. The symbols A1, A2, B1 and B2 denote
the domains of the cross section, described by
Eqs.~(18), (21), (24) and (29),
respectively. The value of $\mu$ corresponds to
$\rho Z/A =1.945\times10^{11}~{\rm g~cm^{-3}}$. The part of the
cross section inside the bar in the left bottom corner
of the top plot is shown separately in Fig.~2.}
\label{F1}
\end{figure}

\begin{figure}
\caption[]{The behaviour of the cross section near the boundary
of the domains B1 and B2. The peaks labeled by the square and
triangle correspond to the transitions $n=2 \to n'=1$ and
$n=1 \to n'=1$, respectively.}
\label{F2}
\end{figure}

\begin{figure}
\caption[]{The influence of the magnetic field on the
NES cross section. The left panel shows
the cross-section as the function of the incident neutrino energy
at $\omega'=15$. The right panel presents the dependence of the cross
section on the energy of the scattered neutrino at $\omega=15$.
Dashed lines show the cross sections at zero magnetic field. The
calculations were done for $\mu=60$ (which corresponds to
$\rho Z/A \simeq 1.9\times 10^{11}~{\rm g~cm^{-3}}$), and $T=3$.}
\label{F3}
\end{figure}
\begin{figure}
\caption[]{The cross section for the case when the scattered neutrino
propagates in the plane ``{\bf k}-{\bf B}" and the angle of
scattering (between ${\bf k}'$ and {\bf k}) is $\theta=90^\circ$.
The scattering cross sections are shown as the functions of
$\omega'/\omega$, at different angles $\vartheta$ between {\bf k}
and {\bf B}. Dashed curves correspond to the zero field limit.}
\label{F4}
\end{figure}
\begin{figure}
\caption[]{The same as in Figure~4, for the case when the scattered
neutrino propagates transverse to the plane ``{\bf k}-{\bf B}".}
\label{F5}
\end{figure}
\begin{figure}
\caption[]{The effect of the replacement
$\vartheta' \leftrightarrow \vartheta$
on the scattering cross section in strong magnetic field.}
\label{F6}
\end{figure}

\end{document}